\documentclass[a4paper,11pt]{article}
\pdfoutput=1 

\usepackage{jcappub} 

\usepackage[T1]{fontenc} 

\usepackage{graphicx}
\usepackage{multirow}
\usepackage{comment}
\usepackage[utf8]{inputenc}
\usepackage[english]{babel}   
\usepackage{float}
\usepackage{amssymb}
\usepackage{amsmath}
\usepackage{slashed}
\usepackage{color}
\usepackage{mathtools}
\usepackage{hyperref}
\usepackage{arydshln}
\usepackage[normalem]{ulem}

\newcommand\nnfootnote[1]{
  \begin{NoHyper}
  \renewcommand\thefootnote{}\footnote{#1}%
  \addtocounter{footnote}{-1}%
  \end{NoHyper}
}
\usepackage{bbold}
\usepackage{soul}
\usepackage{caption}
\usepackage{subcaption}
\usepackage{cancel}
\def\baredth{\;\overline{\raise1.0pt\hbox{$'$}\hskip-6pt
		\partial}\;}
\def\edth{\;\raise1.0pt\hbox{$'$}\hskip-6pt\partial\;}

\newcommand{\litebird}[0]{\textit{L}}
\newcommand{\liteplanck}[0]{\textit{LP}}
\newcommand{\planckso}[0]{\textit{PSO}}
\newcommand{\plancksfour}[0]{\textit{PS4}}
\newcommand{\litesfour}[0]{\textit{LS4}}

\newcommand{\Tstokes}[0]{\textit{T}}
\newcommand{\Qstokes}[0]{\textit{Q}}
\newcommand{\Ustokes}[0]{\textit{U}}
\newcommand{\Vstokes}[0]{\textit{V}}

\newcommand{\Tmodes}[0]{\text{T}}
\newcommand{\Emodes}[0]{\text{E}}
\newcommand{\Bmodes}[0]{\text{B}}
\newcommand{\Vmodes}[0]{\text{V}}

\def\setsymbol#1#2{\expandafter\def\csname #1\endcsname{#2}}
\def\getsymbol#1{\csname #1\endcsname}

\newbox\tablebox    \newdimen\tablewidth
\def\leaderfil{\leaders\hbox to 5pt{\hss.\hss}\hfil}
\def\tablenote#1 #2\par{\begingroup \parindent=0.8em
    \abovedisplayshortskip=0pt\belowdisplayshortskip=0pt
    \noindent
    $$\hss\vbox{\hsize\tablewidth \hangindent=\parindent \hangafter=1 \noindent
    \hbox to \parindent{$^#1$\hss}\strut#2\strut\par}\hss$$
    \endgroup}

\def\L2{\ifmmode L_2\else $L_2$\fi}

\def\DeltaT{\ifmmode \Delta T\else $\Delta T$\fi}
\def\deltat{\ifmmode \Delta t\else $\Delta t$\fi}
\def\fknee{\ifmmode f_{\rm knee}\else $f_{\rm knee}$\fi}
\def\Fmax{\ifmmode F_{\rm max}\else $F_{\rm max}$\fi}
\def\solar{\ifmmode{\rm M}_{\mathord\odot}\else${\rm M}_{\mathord\odot}$\fi}
\def\Msolar{\ifmmode{\rm M}_{\mathord\odot}\else${\rm M}_{\mathord\odot}$\fi}
\def\Lsolar{\ifmmode{\rm L}_{\mathord\odot}\else${\rm L}_{\mathord\odot}$\fi}
\def\inv{\ifmmode^{-1}\else$^{-1}$\fi}
\def\mo{\ifmmode^{-1}\else$^{-1}$\fi}
\def\sup#1{\ifmmode ^{\rm #1}\else $^{\rm #1}$\fi}
\def\expo#1{\ifmmode \times 10^{#1}\else $\times 10^{#1}$\fi}
\def\,{\thinspace}
\def\lsim{\mathrel{\raise .4ex\hbox{\rlap{$<$}\lower 1.2ex\hbox{$\sim$}}}}
\def\gsim{\mathrel{\raise .4ex\hbox{\rlap{$>$}\lower 1.2ex\hbox{$\sim$}}}}

\def\simprop{\mathrel{\raise .4ex\hbox{\rlap{$\propto$}\lower 1.2ex\hbox{$\sim$}}}}
\def\deg{\ifmmode^\circ\else$^\circ$\fi}
\def\pdeg{\ifmmode $\setbox0=\hbox{$^{\circ}$}\rlap{\hskip.11\wd0 .}$^{\circ}
          \else \setbox0=\hbox{$^{\circ}$}\rlap{\hskip.11\wd0 .}$^{\circ}$\fi}
\def\arcs{\ifmmode {^{\scriptstyle\prime\prime}}
          \else $^{\scriptstyle\prime\prime}$\fi}
\def\arcm{\ifmmode {^{\scriptstyle\prime}}
          \else $^{\scriptstyle\prime}$\fi}
\newdimen\sa  \newdimen\sb
\def\parcs{\sa=.07em \sb=.03em
     \ifmmode \hbox{\rlap{.}}^{\scriptstyle\prime\kern -\sb\prime}\hbox{\kern -\sa}
     \else \rlap{.}$^{\scriptstyle\prime\kern -\sb\prime}$\kern -\sa\fi}
\def\parcm{\sa=.08em \sb=.03em
     \ifmmode \hbox{\rlap{.}\kern\sa}^{\scriptstyle\prime}\hbox{\kern-\sb}
     \else \rlap{.}\kern\sa$^{\scriptstyle\prime}$\kern-\sb\fi}
\def\ra[#1 #2 #3.#4]{#1\sup{h}#2\sup{m}#3\sup{s}\llap.#4}
\def\dec[#1 #2 #3.#4]{#1\deg#2\arcm#3\arcs\llap.#4}
\def\deco[#1 #2 #3]{#1\deg#2\arcm#3\arcs}
\def\rra[#1 #2]{#1\sup{h}#2\sup{m}}

\def\dots{\relax\ifmmode \ldots\else $\ldots$\fi}
\def\WHzsr{\ifmmode $W\,Hz\mo\,sr\mo$\else W\,Hz\mo\,sr\mo\fi}
\def\mHz{\ifmmode $\,mHz$\else \,mHz\fi}
\def\GHz{\ifmmode $\,GHz$\else \,GHz\fi}
\def\mKs{\ifmmode $\,mK\,s$^{1/2}\else \,mK\,s$^{1/2}$\fi}
\def\muKs{\ifmmode \,\mu$K\,s$^{1/2}\else \,$\mu$K\,s$^{1/2}$\fi}
\def\muKRJs{\ifmmode \,\mu$K$_{\rm RJ}$\,s$^{1/2}\else \,$\mu$K$_{\rm RJ}$\,s$^{1/2}$\fi}
\def\muKHz{\ifmmode \,\mu$K\,Hz$^{-1/2}\else \,$\mu$K\,Hz$^{-1/2}$\fi}
\def\MJysr{\ifmmode \,$MJy\,sr\mo$\else \,MJy\,sr\mo\fi}
\def\MJysrmK{\ifmmode \,$MJy\,sr\mo$\,mK$_{\rm CMB}\mo\else \,MJy\,sr\mo\,mK$_{\rm CMB}\mo$\fi}
\def\microns{\ifmmode \,\mu$m$\else \,$\mu$m\fi}

\def\muK{\ifmmode \,\mu$K$\else \,$\mu$\hbox{K}\fi}
\def\microK{\ifmmode \,\mu$K$\else \,$\mu$\hbox{K}\fi}
\def\muW{\ifmmode \,\mu$W$\else \,$\mu$\hbox{W}\fi}
\def\kms{\ifmmode $\,km\,s$^{-1}\else \,km\,s$^{-1}$\fi}
\def\kmsMpc{\ifmmode $\,\kms\,Mpc\mo$\else \,\kms\,Mpc\mo\fi}
\providecommand{\sorthelp}[1]{}

\usepackage[dvipsnames]{xcolor}

\def\NHUNIT{\ifmmode {\rm \,cm^{-2}} \else $\rm \,cm^{-2}$ \fi} 

\def\muKcmb{\ifmmode \,\mu$K$_{\rm CMB}$\else \,$\mu$K$_{\rm CMB}$\fi}
\newcommand{\OmegaM}{\ifmmode\Omega_{\rm M}\else $\Omega_{\rm M}$\fi}

\providecommand{\text}[1]{\rm{#1}}

\providecommand{\muK}{\mu\rm{K}}

\newcommand{\begm}{\begin{pmatrix}}
\newcommand{\enm}{\end{pmatrix}}

\def\pmb#1{\setbox0=\hbox{#1}%
    \kern-.025em\copy0\kern-\wd0
    \kern.05em\copy0\kern-\wd0
    \kern-.025em\raise.0433em\box0}

\def\p2Y{\;_2Y}
\def\m2Y{\;_{-2}Y}

\newcommand{\mksym}[1]{\ifmmode {\rm #1}\else #1\fi}

\providecommand{\text}[1]{\rm{#1}}

\providecommand{\muK}{\mu\rm{K}}

\newcommand\ba{\begin{eqnarray}}
\newcommand\ea{\end{eqnarray}}
\newcommand\bea{\begin{eqnarray}}
\newcommand\eea{\end{eqnarray}}

\newcommand\be{\begin{equation}}
\newcommand\ee{\end{equation}}

\title{\boldmath{Unveiling V Modes: Enhancing CMB Sensitivity to BSM Physics with a Non-Ideal Half-Wave Plate}}

\author[a,b]{N. Raffuzzi,}
\author[a,b]{M. Lembo,}
\author[c,a]{S. Giardiello,}
\author[b]{M. Gerbino,}
\author[b]{M. Lattanzi,}
\author[a,b]{P. Natoli,}
\author[a,b,d]{and L. Pagano}

\affiliation[a]{Dipartimento di Fisica e Scienze della Terra, Università degli Studi di Ferrara, via Saragat 1, I-44122 Ferrara, Italy}
\affiliation[b]{Istituto Nazionale di Fisica Nucleare, Sezione di Ferrara, via Saragat 1, I-44122 Ferrara, Italy}
\affiliation[c]{School of Physics and Astronomy, Cardiff University, The Parade, Cardiff, Wales CF24 3AA, United Kingdom}
\affiliation[d]{Institut d'Astrophysique Spatiale, CNRS, Univ. Paris-Sud, Universit\'{e} Paris-Saclay, B\^{a}t. 121, 91405 Orsay cedex, France}

\emailAdd{nicolelia.raffuzzi@unife.it}
\emailAdd{margherita.lembo@unife.it}
\emailAdd{GiardielloS@cardiff.ac.uk}

\abstract{
V-mode polarization of the cosmic microwave background is expected to be vanishingly small in the $\Lambda$CDM model and, hence, usually ignored. Nonetheless, several astrophysical effects, as well as beyond standard model physics could produce it at a detectable level. A realistic half-wave plate - an optical element commonly used in CMB experiments to modulate the polarized signal - can provide sensitivity to V modes without significantly spoiling that to linear polarization. We assess this sensitivity for some new-generation CMB experiments, such as the LiteBIRD satellite, the ground-based Simons Observatory and a CMB-S4-like experiment. We forecast the efficiency of these experiments to constrain the phenomenology of certain classes of BSM models inducing mixing of linear polarization states and generation of \Vmodes\ modes in the CMB. We find that new-generation experiments can improve current limits by 1-to-3 orders of magnitude, depending on the data combination. The inclusion of \Vmodes-mode information dramatically boosts the sensitivity to these BSM models.
}

\begin{document}
\nnfootnote{2010 Mathematics Subject Classification: 05A05, 05A16.} 
\maketitle
\flushbottom

\section{Introduction}
Observations of anisotropies in the cosmic microwave background (CMB) radiation proved a major observational channel for modern cosmology. The Planck satellite observed them in temperature and polarization with unprecedented precision, providing state-of-the-art constraints on cosmology and fundamental physics~\cite{Planck:2018vyg}. Ground-based experiments have complemented these observations, especially targeting the polarization signal~\cite{POLARBEAR:2019kzz, Polarbear:2020lii, ACT:2020gnv, SPT-3G:2021eoc}. 
CMB \Emodes-mode polarization remains a key focus for next-generation experiments, aiming to achieve a cosmic-variance-limited estimate of the optical depth to reionization $\tau$~\cite{LiteBIRD:2022cnt}, the least constrained parameter in the $\Lambda$CDM model. Also, observations of large-scale \Bmodes-mode polarization are crucial for improving sensitivity to, or detecting, the tensor-to-scalar ratio $r$~\cite{LiteBIRD:2022cnt, SimonsObservatory:2018koc, CMB-S4:2020lpa}, ac proxy for primordial gravitational waves. Enhanced sensitivity to small-scale \Emodes-mode and \Bmodes-mode anisotropies will further constrain parameters of $\Lambda$CDM and its extensions~\cite{SimonsObservatory:2018koc, CMB-S4:2020lpa}. Additionally, improved observations of polarized CMB will help explore BSM physics, such as deviations from standard electromagnetism~\cite{Caloni:2022kwp}, neutrino properties~\cite{LiteBIRD:2022cnt}, dark sectors~\cite{Choi:2018gho}, and supersymmetries~\cite{Dalianis:2018afb, PRISM:2013fvg}.

The circular polarization of the CMB, also known as \Vmodes\ modes, is expected to be small in the standard cosmological model, as it is not produced by Thomson scattering during recombination and reionization. Several standard and non-standard physical mechanism can however source a degree of 
circular polarization in CMB photons, by converting linear 
polarization generated at the last scattering surface. Among the standard mechanisms, photon-photon scattering via Heisenberg-Euler interaction at recombination produces the strongest circular polarization~\cite{Motie:2011az, Sawyer:2014maa}. For a comprehensive overview of conventional mechanisms that can source V-mode polarization in the CMB radiation see e.g. Ref.~\cite{Montero-Camacho:2018vgs}. In the realm of non-standard physics, extensions of QED suggest that V-modes arise in the presence of Lorentz-violating operators~\cite{Alexander:2008fp, Colladay:1998fq, Caloni:2022kwp}. Additionally,  Refs.~\cite{Finelli:2008jv, Alexander:2019sqb} show that a pseudo-scalar field or axion inflation leads to V-mode generation. 
Magneto-optic effects are another class of phenomena capable of generating CMB circular polarization\cite{Ejlli:2016avx, Ejlli:2018ucq}.
The detection of circular polarization in the CMB has therefore the potential to offer further evidence for novel physics or more stringent constraints than those available from observations of linear polarization only.

Vast improvements are expected in the observation of the CMB polarized signal both from ground-based experiments, such as the currently running Simons Observatory (SO)~\cite{SimonsObservatory:2018koc} and the next-generation CMB experiment CMB-S4~\cite{CMB-S4:2020lpa}, and from satellites such as the LiteBIRD mission~\cite{LiteBIRD:2022cnt}. Achieving the promised sensitivity to detect the faint cosmological signals hidden in polarization requires monitoring and minimizing systematic effects. Upcoming observations will benefit from improved scanning strategies, including the use of a rotating half-wave plate (HWP) as a polarization modulator for SO and LiteBIRD. Previous studies have already demonstrated the effectiveness of a rotating HWP in mitigating 1/f noise,~\cite{Johnson:2006jk}, and reducing temperature-to-polarization leakage resulting from pair-differencing of orthogonal detectors~\cite{Bryan:2010eh, ABS:2016rpo}. However, despite the addition of a HWP is remarkably useful, the inclusion of another optical element in the acquisition chain brings along additional systematic effects, which in turn may compromise the final scientific product. One example is a HWP with a non-ideal phase shift, i.e., a HWP inducing a phase difference $\neq 180^\circ$ between the two components of the electromagnetic wave traversing it. Interestingly enough, this deviation, possibly degrading the sensitivity to linear polarization, can also cause a coupling between total intensity and circular polarization. This would allow to investigate the presence of \Vmodes\ modes in the CMB radiation~\cite{SPIDER:2017kwu}. In this paper, we exploit this possibility showing how forthcoming CMB experiments, equipped with HWPs, can be sensitive to circular polarization and, in turn, provide valuable information on a specific class of BSM physical models. We investigate a simple mechanism named the Generalized Faraday Effect (GFE), i.e. the mixing of CMB polarization states, including a partial conversion of linear into circular polarization~\cite{Kennett:1998bal,Lembo:2020ufn}.

The paper is structured as follows: Section~\ref{section: Sensitivity to V-modes} introduces the mathematical formalism (Jones calculus~\cite{Jones:1941rya}) describing the optical behaviour of a (realistic) HWP, the instrumental systematic effect induced by a non-ideal phase shift, as well as the expected sensitivity to \Vmodes\ modes; Section~\ref{section: Theory} illustrates the specific mechanism generating circular polarization and its cosmological phenomenology, while Section~\ref{section: Methodology} describes the analysis method adopted in this work; in Section~\ref{section: Results}, we present the results of the analysis, forecasting the performance of future CMB experiments. Finally, Section~\ref{section: Conclusions} provides conclusions. Appendix~\ref{section: Appendix for triangular plots} includes a full set of plots with posterior probability distributions of the key cosmological parameters for this work and for each experiment configuration considered in this analysis.

\section{Sensitivity to \Vmodes\ modes} \label{section: Sensitivity to V-modes}
In this section, we give an overview of the mathematical framework to model a HWP with non-ideal phase shift $\beta$ (for a detailed description, see, e.g., \cite{ODea:2006tvb,Giardiello:2021uxq}). We also provide an estimate of the sensitivity to \Vmodes\ modes for a generic CMB experiment employing a non-ideal HWP.

\subsection{The non-ideal HWP}
An ideal half-wave plate (HWP) is an optical device that induces a $\pi$-phase shift between the two orthogonal components of the incident wave. In the Jones formalism, the matrix describing the behavior of an ideal HWP is:
\begin{equation*}
    \mathbf{J}_{\mathrm{HWP}}^{Ideal} =
    \begin{pmatrix}
        1 & 0\\
        0 & -1
    \end{pmatrix}.
\end{equation*}
A realistic HWP can be described by the more general Jones matrix \cite{ODea:2006tvb}:
\begin{equation} \label{eq: non-ideal hwp}
    \mathbf{J}_{\mathrm{HWP}} = 
    \begin{pmatrix}
        1+h_1 & \zeta_1 e^{i \chi_1}\\
        \zeta_2 e^{i \chi_2} & -(1+h_2) e^{i \beta}
    \end{pmatrix},
\end{equation}
where all but one entries of the matrix (e.g., $J_{11}$) are complex.
The parameters $h_{1,2}$ are real and negative-defined, and indicate deviations from unitary transmission of the two polarized components, $E_{x,y}$, of the incident light due to absorption. The parameters $\zeta_{1,2}$ and $\chi_{1,2}$ correspond to the amplitudes and phases of the off-diagonal terms that are responsible for cross-polarization, i.e., mixing between the two orthogonal components of the wave. The parameter $\beta$ represents the departure from the ideal phase shift of $\pi$. In this work, we focus on the effects of a non-vanishing $\beta$ and how it can be used to measure circular polarization. In the following, we assume for simplicity that the effects due to absorption and cross-polarization are negligible and fix the relative parameters in Eq.~\ref{eq: non-ideal hwp} to zero. The Jones matrix of a HWP with non-ideal phase shift and spinning with angular velocity $\omega$ is:
\begin{equation}
    \mathbf{J}_{\mathrm{HWP}}(\theta)=\mathbf{R}^{\mathrm{T}}(\theta) \mathbf{J}_{\mathrm{HWP}} \mathbf{R}(\theta)=\left(\begin{array}{cc}
\mathrm{J}_{11}(\theta) & \mathrm{J}_{12}(\theta) \\
\mathrm{J}_{21}(\theta) & \mathrm{J}_{22}(\theta)
\end{array}\right), \quad \text {with} \quad \mathbf{R}(\theta)=\left(\begin{array}{cc}
\cos \theta & \sin \theta \\
-\sin \theta & \cos \theta
\end{array}\right),
\end{equation}
where $\theta=\omega t$ is the time-dependent angle between the HWP fast axis and the x-axis in the chosen reference frame, and:
\begin{equation}
    \begin{aligned}
&\mathrm{J}_{11}(\theta)=\cos^2\theta-\text{e}^{i \beta} \sin^2\theta \\
&\mathrm{J}_{12}(\theta)=J_{21}(\theta)=\left(1+\text{e}^{i \beta}\right) \cos\theta \sin\theta \\
&\mathrm{J}_{22}(\theta)=-\text{e}^{i \beta} \cos^2\theta+\sin^2\theta.
\end{aligned}
\end{equation}

In the case under study, the complete optical chain traversed by the incoming signal includes a rotating HWP, two orthogonal polarizers, and the detector. Eventually, the Jones matrix of the full optical chain\footnote{
Note that this optical chain does not include information about the instrument orientation. The primary goal of this analysis is not to define an explicit scanning strategy, but rather to establish the polarimeter's efficiency through variations in the HWP phase-shift. See \cite{Giardiello:2021uxq} for details.}
is:
\begin{equation}
    \mathbf{J}_{x,y}(\theta)=\mathbf{J}_{\mathrm{pol},(x,y)} \mathbf{R}^{T}(\theta) \mathbf{J}_{\mathrm{HWP}} \mathbf{R}(\theta),
    \quad \text {with} \quad \mathbf{J}_{\mathrm{pol,x}}=\left(\begin{array}{cc}
        1 & 0 \\
        0 & 0
        \end{array}\right), \quad
    \mathbf{J}_{\mathrm{pol,y}}=\left(\begin{array}{cc}
        0 & 0 \\
        0 & 1
        \end{array}\right).
\end{equation}

So far, we have considered a scenario where the incoming wave is fully polarized and has a quasi-monochromatic frequency. However, the CMB signal is only partially polarized, and a CMB experiment measures the time-averaged incident intensity. Therefore, it is more natural to express the signal in terms of the Stokes vector $s = (\Tstokes,\Qstokes,\Ustokes,\Vstokes)$, and employ the Müller formalism~\cite{Jones:2006ac}, which also simplifies tracking the impact on each Stokes parameter. 

Given a Jones matrix, it is straightforward to compute the corresponding $4\times4$ Müller matrix. Explicitly, the Müller elements, for the x-oriented polarizer, are:
\begin{equation} \label{eq: Müller terms x}
\begin{aligned}
    \mathbf{M}_{\Tstokes \Tstokes}^{x} &= \frac{1}{2} \left (\left |\mathrm{J}_{11} \right|^{2} + \left |\mathrm{J}_{12} \right|^{2} \right) =
    \frac{1}{2}\\
    \mathbf{M}_{\Tstokes \Qstokes}^{x} &= \frac{1}{2} \left (\left |\mathrm{J}_{11} \right|^{2} - \left |\mathrm{J}_{12} \right|^{2} \right) =
    \frac{1}{2} \operatorname{Cos} \left( \frac{\beta}{2}\right)^{2} \operatorname{Cos}(4 \theta)\\
    \mathbf{M}_{\Tstokes \Ustokes}^{x} &= \operatorname{Re}\left[\left(\mathrm{J}_{11} \mathrm{J}_{12}^{*}\right)\right] =
    \frac{1}{2} \operatorname{Cos} \left( \frac{\beta}{2} \right)^2 \operatorname{Sin}(4 \theta)\\
    \mathbf{M}_{\Tstokes \Vstokes}^{x} &= \operatorname{Im}\left[\left(\mathrm{J}_{11} \mathrm{J}_{12}^{*}\right)\right] =
    -\frac{1}{2} \operatorname{Sin} (\beta) \operatorname{Sin} (2 \theta).
\end{aligned}
\end{equation}
The y-oriented elements can be obtained by applying a 90-degree rotation ($4\theta\rightarrow 4\theta+\pi/2$) to the above expressions.
Finally, the total power collected by the detector is:
\begin{equation} \label{eq:bolometer}
    d_{\mathrm{obs}} = \Sigma_{i=x,y} \left(
    \mathrm{M}_{\Tstokes \Tstokes}^{i} \; \Tstokes +
    \mathrm{M}_{\Tstokes \Qstokes}^{i} \; \Qstokes +
    \mathrm{M}_{\Tstokes \Ustokes}^{i} \; \Ustokes +
    \mathrm{M}_{\Tstokes \Vstokes}^{i} \; \Vstokes
    \right).
\end{equation}

From Eqs.~\ref{eq: Müller terms x} - \ref{eq:bolometer}, it is clear that, when $\beta=0$, there is no contribution to the observed power from a non-vanishing V-mode signal. Conversely, the departure of the phase shift from the ideal value of $\pi$ offers a handle to the detection of \Vmodes\ modes. 

\subsection{Sensitivity to Stokes parameters}
We now quantify the sensitivity to \Vmodes\ modes of a CMB experiment equipped with a HWP with non-ideal phase shift. For the sake of simplicity, we consider Gaussian, stationary and uncorrelated instrumental noise of variance $\sigma^2_{\mathrm{pix}}$ in real space. This simplifies the expression of the noise covariance matrix in pixel space\footnote{Equation~\ref{eq:bolometer} can be generalized in matrix form as $\mathbf{d}_\mathrm{obs}=\mathbf{A\,m+n}$, where $\mathbf{d,\,n}$ are vectors of the data and noise of $N_t$ time samples, $\mathbf{m}$ is a vector of $4 N_\mathrm{pix}$ Stokes parameters describing the sky signal in pixel space and $\mathbf{A}$ is a sparse $N_t \times 4 N_\mathrm{pix}$ pointing matrix mapping how each pixel in the sky is converted in a time sample according to Eq.~\ref{eq:bolometer}. The general linear solution to the matrix equation for each pixel $\textbf{pix}$ is given by:
\begin{equation} \label{eq: mapmaking}
    \hat{\mathbf{m}}_{\mathrm{pix}} = (\mathbf{A}^\mathrm{T}\mathbf{N}^{-1} \mathbf{A})^{-1} _{\mathrm{pix}} \: (\mathbf{A}^\mathrm{T} \mathbf{N}^{-1}\mathrm{\mathbf{d}})_{\mathrm{pix}},
\end{equation} \nonumber
where 
$\mathbf{N}=\langle \mathbf{n\,n}^T\rangle$ is the noise covariance matrix in time domain and $(\mathbf{A}^\mathrm{T}\mathbf{N}^{-1} \mathbf{A})_{\mathrm{pix}}^{-1}$ is the noise covariance matrix in pixel space. For the simple noise properties considered in this work, i.e. $\mathbf{N}=\mathbf{diag}(\sigma^2)$, the noise covariance matrix in a pixel is simply given by $\sigma^2_\mathrm{pix}=\sigma^2\left(\mathbf{A}^\mathrm{T} \mathbf{A}\right)_{\mathrm{pix}}^{-1}$. 
This becomes $\sigma^2_\mathrm{pix} =  \frac{4 \sigma^2}{N_\mathrm{hit}} \mathbf{diag}\left(1,\,2/\cos(\beta/2)^4,\,2/\cos(\beta/2)^4,\,2/\sin(\beta)^2\right)$, where $N_{hit}$ is the number of hits on that pixel, since we assume a very homogeneous and redundant scanning strategy. See, e.g.~\cite{Giardiello:2021uxq} and references therein for further details.
}.

The square root of the diagonal elements of the inverse noise covariance in pixel space provides an estimate of the experimental noise per pixel to each Stokes parameter. In the case of a HWP with non-ideal phase shift, the noise -- which we normalize to the total intensity \Tstokes{} one -- is a function of $\beta$:
\begin{equation} \label{eq: sigma of betas}
\begin{split}
    \left(\frac{\sigma_{\Qstokes,\Ustokes} \: (\beta)}{\sigma_{\Tstokes}}\right)_\mathrm{pix} &= \frac{\sqrt{2}}{\operatorname{Cos}^2(\beta /2)} \\
    \left(\frac{\sigma_{\Vstokes} \: (\beta)}{\sigma_{\Tstokes}}\right)_\mathrm{pix} &= \frac{\sqrt{2}}{\operatorname{Sin}(\beta)}.
\end{split}
\end{equation}

\begin{figure}[ht]
\centering
\includegraphics[width=0.9\linewidth]{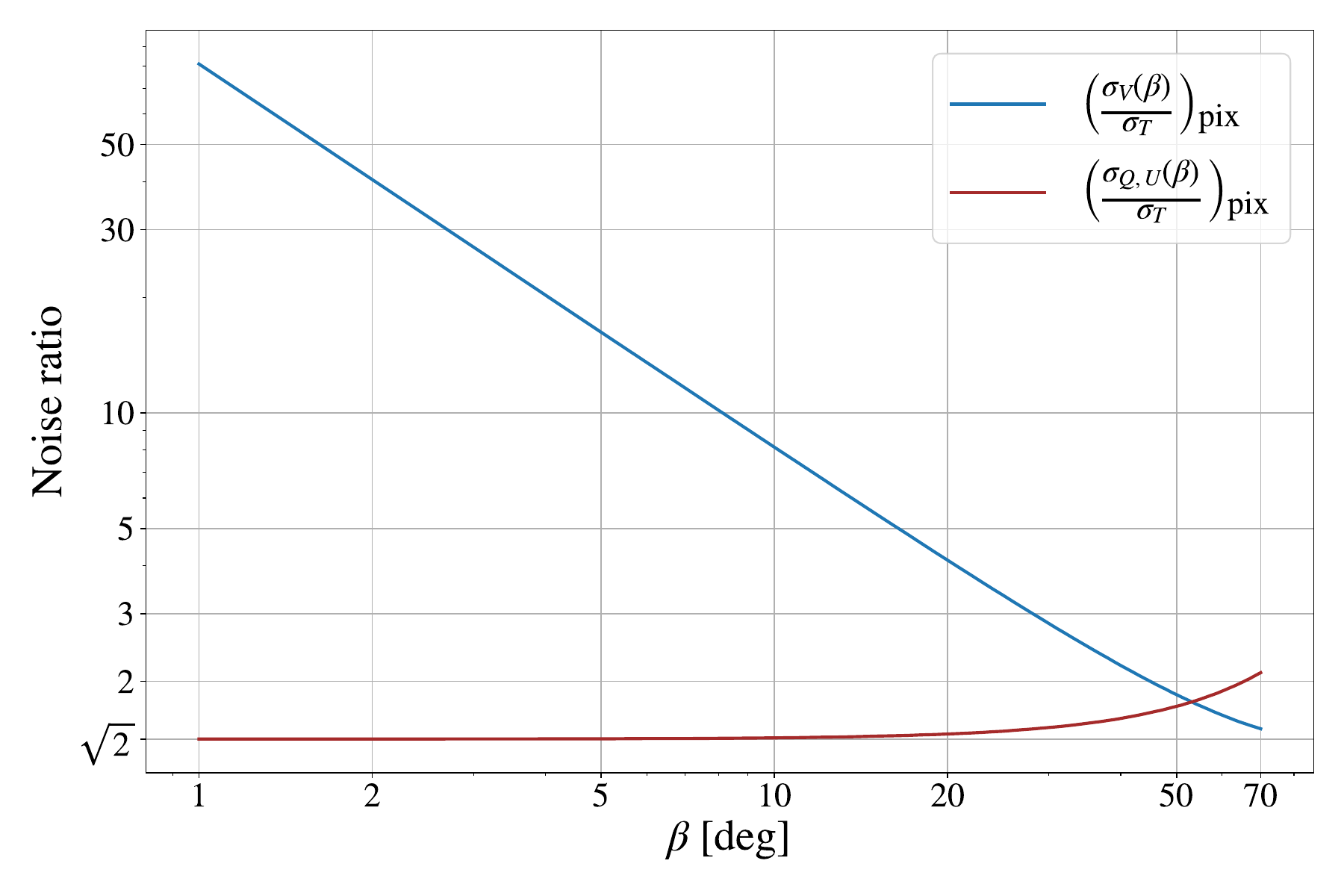}
\vspace*{-1mm}
\caption{Noise ratio of the Stokes parameters for linear polarization $\sigma_{\Qstokes, \, \Ustokes, \, \mathrm{pix}}(\beta)$ (red) and circular polarization $\sigma_{\Vstokes, \, \mathrm{pix}}(\beta)$ (blue), normalized to the total intensity noise, as a function of the HWP non-ideal phase shift $\beta$. Typical values for $\beta$ range from a few degrees to a few tenths \cite{Pisano:2020qxq, Pisano:06, Savini:06}. 
For ideal CMB experiments, $(\sigma_{Q,\,U}/\sigma_\Tstokes)_{\mathrm{pix}} = \sqrt{2}$ for polarization-sensitive detectors.
In presence of a non-ideal phase shift, the noise level for circular polarization $\sigma_{\Vstokes,  \, \mathrm{pix}}(\beta)$ decreases rapidly with increasing $\beta$, while the noise level for linear polarization $\sigma_{\Qstokes, \, \Ustokes,  \, \mathrm{pix}}(\beta)$ does not increase significantly until very large values of $\beta$ are reached. As a result, a slightly non-ideal HWP (small value of $\beta$) can help gain sensitivity to \Vmodes\ modes without significantly degrading the sensitivity to linear polarization.}
\label{fig:sensitivity vs beta}
\end{figure}

In Figure~\ref{fig:sensitivity vs beta}, we show the sensitivity to the Stokes parameters \Qstokes{}, \Ustokes{}, \Vstokes{}, normalized to the sensitivity to the Stokes \Tstokes{}, as a function of the phase shift $\beta$. The expected noise ratio for ideal CMB experiments is $(\sigma_{\Qstokes,\Ustokes}/\sigma_{\Tstokes})_{\mathrm{pix}}=\sqrt{2}$. The non-ideal phase shift leads to a mixing of power among the \Qstokes, \Ustokes, \Vstokes{} Stokes parameters. This can be understood from the fact that the sensitivity to \Qstokes{} and \Ustokes{} decreases (i.e., the noise ratio increases, in red in Fig.~\ref{fig:sensitivity vs beta}) while that of \Vstokes{} (blue) increases for larger values of $\beta$. However, while the sensitivity to \Vstokes{} rapidly changes with $\beta$, the degradation of the sensitivity to \Qstokes, \Ustokes{} significantly deviates from the ideal case only for fairly large values of $\beta\gtrsim 20^{\circ}$ (for comparison, typical values for $\beta$ vary from a few degrees up to a few tenths \cite{Pisano:2020qxq, Pisano:06, Savini:06}).

It is also instructive to work out the sensitivity to \Vmodes\ modes in harmonic space for a given experimental setup, as described in terms of the linear polarization noise level $\sigma_Q$, angular resolution $\theta_{\mathrm{FWHM}}$, observed sky fraction $f_\mathrm{sky}$ and non-ideality HWP parameter $\beta$. Note that in the rest of this section we will denote with $\sigma_P$ the nominal linear polarization noise, corresponding to an ideal HWP plate, i.e. $\sigma_Q=\sigma_Q(\beta = 0) = \sqrt{2} \sigma_T$. In this way factors of $\sin\beta$ and $\cos\beta$ will appear explicitly in the equations.

The observed power spectrum of the signal $\Vmodes$ modes $\hat C_\ell^{\Vmodes \Vmodes}$ has an associated variance $\Sigma^2_\ell$:
\begin{equation}
    \Sigma^2_{\ell} = \frac{2}{(2\ell +1)f_{sky}} \; \left[ C_\ell^{\Vmodes \Vmodes,\mathrm{fid}} +\frac{ \mathrm N_\ell}{\mathrm B^2 _\ell} \frac{1}{\operatorname{Sin}^2(\beta)} \right],
\end{equation}
where $C_\ell^{\Vmodes \Vmodes,\mathrm{fid}}$ is the true underlying power spectrum of the signal, $N_\ell$ is the noise power spectrum of the linear polarization, assuming an ideal half-wave plate, and $B_\ell$ is the harmonic equivalent of a Gaussian beam of width $\theta_{\mathrm{FWHM}}$. We assume a white noise power spectrum $N_\ell = \sigma_Q^2$, and use a fiducial Gaussian approximation for the likelihood ${\mathcal L}(C_\ell^{\Vmodes \Vmodes}) = \mathrm{Pr}(\hat C_\ell^{\Vmodes \Vmodes}|C_\ell^{\Vmodes \Vmodes})$ of a theoretical power spectrum $C_\ell^{\Vmodes \Vmodes}$:
\begin{equation}
    - 2 \ln {\mathcal L}(C_\ell^{\Vmodes \Vmodes}) = \frac{(C_\ell^{\Vmodes \Vmodes}-\hat C_\ell^{\Vmodes \Vmodes})^2}{\Sigma^2_\ell},
\end{equation}
For the purpose of our analysis, we take $\hat C_\ell^{\Vmodes \Vmodes} = \langle \hat C_\ell^{\Vmodes \Vmodes} \rangle = C_\ell^{\Vmodes \Vmodes,\mathrm{fid}}$. The inverse probability $\mathrm{Pr}(C_\ell^{\Vmodes \Vmodes}|\hat C_\ell^{\Vmodes \Vmodes})$ is computed through the use of Bayes' theorem with a flat prior on $C\ell$.

\begin{figure}[ht]
\centering
\includegraphics[width=0.9\linewidth]{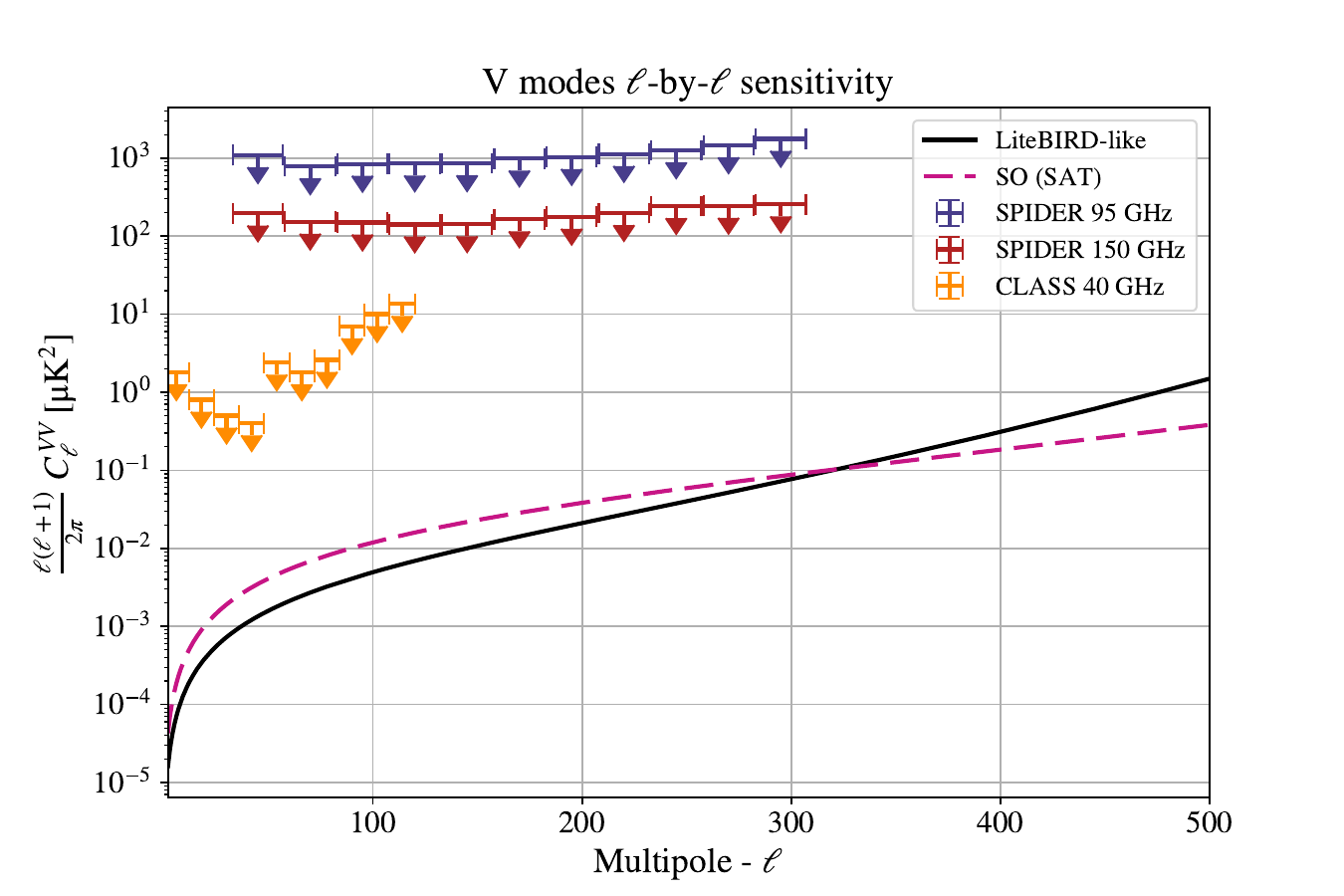}
\vspace*{0mm}
\caption{The figure illustrates the $\ell$-by-$\ell$ (i.e. no binning) 95\% C.L. sensitivity for current and future CMB experiments to a \Vmodes\Vmodes-spectrum. Assuming a non-ideal HWP with a phase shift of $\beta=10^\circ$, the black curve represents the sensitivity for a LiteBIRD-like experiment and the dark magenta curve corresponds to the sensitivity for a SO SAT experiment. Details on the full experimental setup adopted to obtain these curves are in Tab.~\ref{table: noise list}. For comparison, the current bounds on \Vmodes\ modes at different frequencies and $\ell$-ranges from the balloon-borne SPIDER~\cite{SPIDER:2017kwu} (blue and red) and the ground-based CLASS~\cite{Padilla:2019dhz} (yellow) are also shown.
}
\label{fig: 1D sensitivity vs CLASS and SPIDER}
\end{figure}

Armed with the above, we compute, for each multipole $\ell$, the 95\% upper limit on $C_\ell^{\Vmodes \Vmodes}$ in the assumption of vanishing V-mode signal, i.e., for $C_\ell^{\Vmodes \Vmodes,\mathrm{fid}}=0$, and for different experimental configurations. The results are shown in Fig.~\ref{fig: 1D sensitivity vs CLASS and SPIDER}. Here, we show the results for a LiteBIRD-like satellite (black solid curve) and for a ground-based Simons Observatory/SO SAT\footnote{Only the Small Aperture Telescopes (SATs) are equipped with a HWP.} experiment (magenta dashed line), assuming $\beta=10^\circ$ in both cases. The other parameters describing each configuration can be read in Tabs.~\ref{table: noise list} (noise and angular resolution) and~\ref{Table: experimental configuration} (sky coverage).
For comparison, the current 95\% upper bounds from the balloon-borne experiment SPIDER~\cite{SPIDER:2017kwu} (blue and red) and the ground-based CLASS~\cite{Padilla:2019dhz} (yellow) are also shown. A LiteBIRD-like experiment would improve current bounds by several orders of magnitude and extend the sensitivity over a broader range of angular scales. On the other hand, a ground-based experiment like SO will dominate at smaller scales. 

\begin{figure}[ht]
\centering
\includegraphics[width=0.9\linewidth]{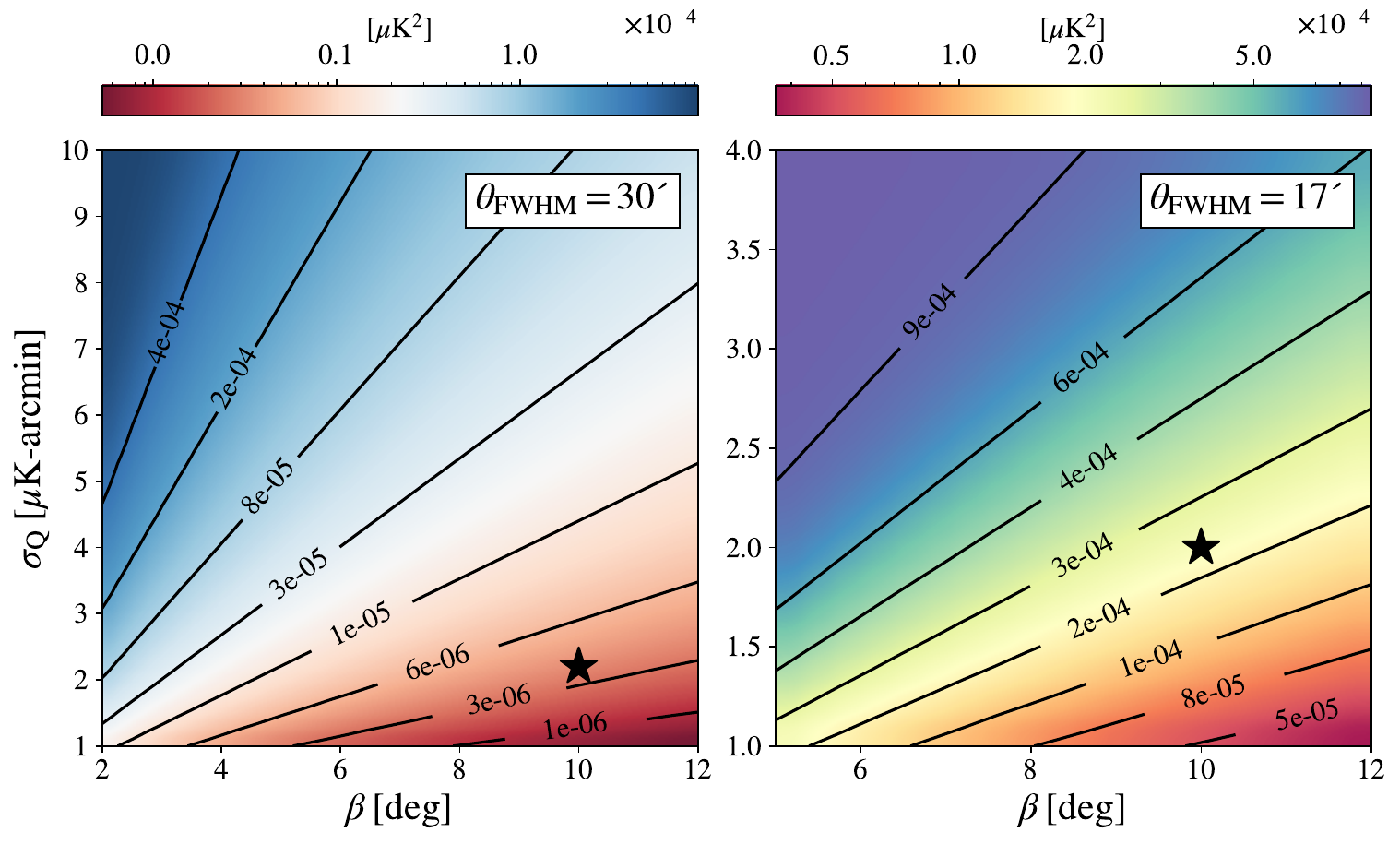}
\vspace*{-1mm}
\caption{
    Contours of constant V-mode 95\% sensitivity for a scale-invariant \Vmodes\Vmodes\ spectrum as function of non-ideal phase-shift $\beta$ and nominal (i.e., assuming $\beta=0$) linear polarization sensitivity $\sigma_{\rm Q}$. Left panel: the sensitivity is integrated over the multipole range $2 \leq \ell \leq 300$ for an angular resolution of $30$ arcmin, similar to that achievable by LiteBIRD from space. For the LiteBIRD experiment, the expected noise level is $\sigma_Q \simeq 2.2 \:\mu \mathrm{K\,arcmin}$, integrated over 22 frequency channels and 3-year observation time~\cite{LiteBIRD:2022cnt}. The star marks the combination of this  noise level with a value of the nonideality parameter $\beta=10^\circ$ (chosen arbitrarily).  Right panel: the sensitivity is integrated over $50 \leq \ell \leq 300$ for an angular resolution of $17$ arcmin, similar to that achievable by the ground-based SO-SAT experiment. For the SO SAT, the noise level is $\sigma_Q \simeq 2 \:\mu \mathrm{K\,arcmin}$, combining the 93 and 145 GHz bands~\cite{SimonsObservatory:2018koc}. The star marks the combination of this  noise level with a value of the nonideality parameter $\beta=10^\circ$ (chosen arbitrarily). 
    }
\label{fig: 2D sensitivity vs beta and sigma Q/U}
\end{figure}

In Figure~\ref{fig: 2D sensitivity vs beta and sigma Q/U}, we show iso-contours of constant integrated 95\% sensitivity to a scale-invariant \Vmodes\Vmodes\ spectrum,  as a function of $\beta$ and the instrumental noise level in linear polarization, $\sigma_\Qstokes$. In the left (right) panel, the angular resolution and sky fraction are fixed to the values corresponding the LiteBIRD (SO) experiment. The two panels also correspond to different choices of the observed multipole range, again representatives of the aforementioned experiments: $2\leq\ell\leq300$ in the left panel and $50\leq\ell\leq300$ in the right panel. As expected, the sensitivity increases for smaller values of $\sigma_{\Qstokes}$ and large values of $\beta$. The stars in the plot indicate the expected performance of a LiteBIRD-like (left panel) and SO (right panel) experiment for  $\beta=10^\circ$.\par
    
\section{Production of V modes} \label{section: Theory}
In the previous section, we have quantified the sensitivity to \Vmodes\ modes assuming a generic scale-invariant spectrum. In the following, we work out the sensitivity of different combinations of CMB experiments to a \Vmodes-mode signal generated in the phenomenological framework studied in~\cite{Lembo:2020ufn}. We now briefly recall the basis of this formalism.\par

We consider a simple class of models allowing for the in vacuo mixing of CMB polarization states during propagation, including the conversion of linear into circular polarization. The mixing of polarization states in a weakly anisotropic medium is dubbed Generalized Faraday Effect~(see sec. 14.3 of \cite{Melrose:1991eu} for a description of the general framework, and \cite{Kennett:1998bal} for an application to extragalactic synchrotron sources), or GFE. As the name suggests, the usual Faraday rotation is a particular case of GFE.
The radiative transfer equation for the linearly polarized radiation propagating in a weakly anisotropic non-absorbing medium reads \cite{Melrose:1991eu}
\begin{equation} \label{eq:transfer eq}
    \frac{\mathrm{d}}{\mathrm{d} s} \boldsymbol{S}=\boldsymbol{\rho} \times \boldsymbol{S}
\end{equation}
where $s$ is an affine parameter following the direction of wave propagation, and $\boldsymbol{S} = (Q,U,V)$ is the polarization Stokes vector, precessing about the direction of $\boldsymbol{\rho} = (\rho_Q,\rho_U,\rho_V)$, which is determined by the dielectric tensor of the medium and thus encodes its optical properties.
If $\hat{\boldsymbol{\rho}}\equiv\boldsymbol{\rho}/|\boldsymbol{\rho}|$ is directed along the $V$-axis in the polarization space ($\rho_Q=\rho_U=0$), then the Stokes vector $\boldsymbol{S}$ precesses about it. The linear polarization component, i.e. the projection of $S$ on the $(Q,\,U)$ plane, will thus rotate while mantaining a constant magnitude. This is the usual Faraday rotation, where no circular polarization is generated, and only the $Q$ and $U$ states mix among themselves.
On the other hand, if ${\boldsymbol{\rho}}$ has non-vanishing $Q$ and $U$ components, a partial conversion of linear into circular polarization occurs, generating non-zero \Vmodes\ modes. 

In Ref. \cite{Lembo:2020ufn}, some of us have developed a formalism that allows to derive the modifications to the CMB spectra in temperature, linear and circular polarization, and their cross-correlations, given an effective dielectric tensor that describes the optical properties of the space traversed by CMB photons. This allows to be agnostic about the specific model beyond the mixing among CMB polarization components and just capture the basic phenomenology. Therefore, once the phenomenological parameters are constrained, it is possible to interpret the results in light of a specific physical model that could have induced GFE; see, e.g.~\cite{Caloni:2022kwp} for a recent application.

Here we consider the case of an effective dieletric tensor that is homogeneous and independent of both the direction and magnitude of the radiation wave vector.

In this specific case, parity-violating CMB cross spectra are vanishing, $\mathrm{TB}=\mathrm{EB}=\mathrm{EV}=\mathrm{BV}=0$ (see Ref.~\cite{Lembo:2020ufn} for details).

The remaining CMB power spectra $C_\ell^{XY}$ (with $X,\,Y=T,\,E,\,B,\,V$) can be expressed in terms of the CMB spectra produced at the last-scattering surface $\widetilde{C}_\ell^{XY}$:
\begin{equation}\label{eq: cmb power spectra}
    \begin{aligned}
&C_{\ell}^{\Tmodes \Emodes}=\left(1-\frac{\beta_V^2+\beta_E^2}{8 \pi}\right) \widetilde{C}_{\ell}^{\Tmodes \Emodes} \\
&C_{\ell}^{\Emodes \Emodes}=\left(1-\frac{\beta_V^2+\beta_E^2}{4 \pi}\right) \widetilde{C}_{\ell}^{\Emodes \Emodes}+\frac{\beta_V^2}{4 \pi}\left[\mathcal{W}_{\ell}^{(1)} \widetilde{C}_{\ell}^{\Emodes \Emodes}+\mathcal{W}_{\ell+1}^{(1)} \widetilde{C}_{\ell+1}^{\Bmodes \Bmodes}+\mathcal{W}_{\ell-1}^{(1)} \widetilde{C}_{\ell-1}^{\Bmodes \Bmodes}\right] \\
&C_{\ell}^{\Bmodes \Bmodes}=\left(1-\frac{\beta_V^2+\beta_E^2}{4 \pi}\right) \widetilde{C}_{\ell}^{\Bmodes \Bmodes}+\frac{\beta_V^2}{4 \pi}\left[\mathcal{W}_{\ell}^{(1)} \widetilde{C}_{\ell}^{\Bmodes \Bmodes}+\mathcal{W}_{\ell+1}^{(1)} \widetilde{C}_{\ell+1}^{\Emodes \Emodes}+\mathcal{W}_{\ell-1}^{(1)} \widetilde{C}_{\ell-1}^{\Emodes \Emodes}\right] \\
&C_{\ell}^{\Vmodes \Vmodes}=\frac{\beta_E^2}{\pi}\left[\mathcal{W}_{\ell+2}^{(2)} \widetilde{C}_{\ell+2}^{\Bmodes \Bmodes}+\mathcal{W}_{\ell+1}^{(2)} \widetilde{C}_{\ell+1}^{\Emodes \Emodes}+\mathcal{W}_{\ell}^{(2)} \widetilde{C}_{\ell}^{\Bmodes \Bmodes}+\mathcal{W}_{\ell-1}^{(2)} \widetilde{C}_{\ell-1}^{\Emodes \Emodes}+\mathcal{W}_{\ell-2}^{(2)} \widetilde{C}_{\ell-2}^{\Bmodes \Bmodes}\right],
    \end{aligned}
\end{equation}
where $\mathcal{W} ^{(1,2)}$ are combinations of Wigner $3j$-symbols, and the phenomenological parameters $\beta_E^2$ and $\beta_V^2$ are, respectively, combinations of the diagonal and off-diagonal components of the effective susceptibility tensor \cite{Lembo:2020ufn}.

\begin{figure}[h]
\centering
\includegraphics[width=15.6cm,height=10cm,keepaspectratio]{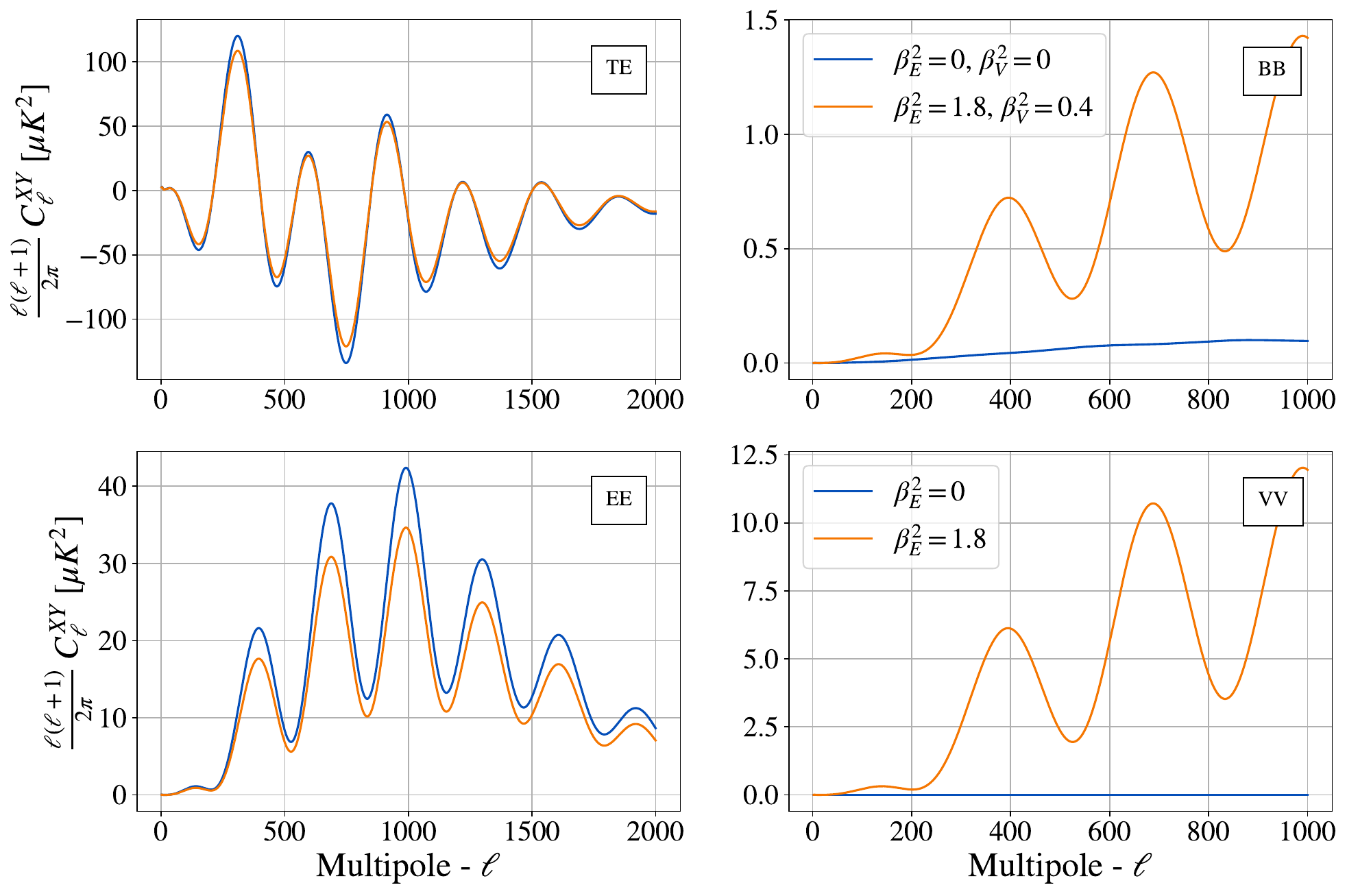}
\vspace*{0mm}
\caption{Impact of Generalized Faraday Effect (GFE) on polarization power spectra. The reference (blue) power spectra are computed using Planck best-fit values from \Tmodes\Tmodes, \Tmodes\Emodes, \Emodes\Emodes, low\Emodes{} + lensing~\cite{Planck:2018vyg} and vanishing GFE parameters, as reported in Tab.~\ref{Table: fiducial values}, while the modified spectra (orange) are calculated fixing $\beta_V^2=0.4$ and $\beta_E^2=1.8$ (chosen arbitrarily to make the GFE visible).
}
\label{fig: faraday rotated power spectra}
\end{figure}

From Eqs.~\ref{eq: cmb power spectra} and Fig.~\ref{fig: faraday rotated power spectra}, it is clear that the modification of the CMB polarization signal due to this particular realization of GFE has three main effects:
\begin{itemize}
    \item rescaling of the amplitude of $\Tmodes\Emodes,\,\Emodes\Emodes,\,\Bmodes\Bmodes$ spectra via the combination $\beta_V^2+\beta_E^2$;
    \item mixing between \Emodes{} and \Bmodes\ modes (cosmic birefringence), proportional to $\beta_V^2$;
    \item sourcing of \Vmodes\ modes from \Emodes{} and \Bmodes\ modes, proportional to $\beta_E^2$.
\end{itemize}

\section{Analisys method and datasets} \label{section: Methodology}
In this section, we present methodology and data employed in the likelihood analysis.
We simulate the performance of different upcoming CMB experiments (satellite and ground-based) and forecast their sensitivity to the phenomenological GFE parameters $\beta_{E,V}^2$ introduced in the previous section. To this scope, we perform a Monte Carlo Markov Chain (MCMC) analysis to obtain forecasted constraints on $\beta_E^2$ and $\beta_V^2$ jointly with the six $\Lambda$CDM parameters and the tensor-to-scalar ratio $r$\footnote{The full set of sampled parameters is as follows: the angular scale of the sound horizon at recombination, $\theta$, the cold dark matter and baryon physical densities today, $\omega_{cdm}$ and $\omega_b$ respectively, the amplitude and spectral index of the spectrum of primordial scalar perturbations, $\log(10^{10}A_s)$ and $n_s$ respectively, the optical depth to reionization $\tau$, the tensor-to-scalar ratio $r$ and the GFE phenomenological parameters $\beta_E^2$ and $\beta_V^2$. We assume adiabatic initial conditions and consider inflation consistency, which allows to express the spectral index of tensor perturbations as $n_t=-r/8$. We consider one massive neutrino of mass $m_\nu=0.06\,\mathrm{eV}$ and two massless neutrinos, for a total number of relativistic degrees of freedom at recombination of $N_\mathrm{eff}=3.044$.}. We use a customized version\footnote{The code is publicly available at this link: \url{https://github.com/nraffuzz/CosmoMC_GFE}} of the \texttt{CosmoMC}~\cite{PhysRevD.66.103511} package to perform the analysis. We assume convergence of the MCMC chain by requiring that the Gelman-Rubin statistics $R-1\sim0.01$. 

We run forecasts for five combinations of data sets which simulate the performance of different CMB experiments:
\begin{itemize}
    \item a future LiteBIRD-like satellite experiment. We refer to this case as ``\litebird'';
    \item LiteBIRD-like in combination with a simulated version of the Planck data, labeled ``\liteplanck'';
    \item a currently running Simons Observatory (SO) ground-based experiment, in combination with Planck, labeled ``\planckso''. We include simulated performance of the observations with both the Large Aperture Telescope (LAT) which targets the smallest angular scales and the array of Small Aperture Telescopes (SATs) which target \Bmodes\ modes at degree angular scales. The HWP is deployed on SATs only;
    \item Planck in combination with a next-generation ground-based experiment, i.e. CMB-S4-like, labeled ``\plancksfour''. As for SO, we include information from both LAT and SATs. We note that the baseline design of CMB-S4, as described in \cite{Abazajian:2019eic,CMB-S4:2020lpa}, does not include a HWP. Therefore, when referring to a CMB-S4-like experiment, we actually refer to the combination of expected sensitivity as determined by the sky coverage, angular resolution and noise properties reported in \cite{Abazajian:2019eic,CMB-S4:2020lpa}, augmented with the use of a HWP in the telescopes targeting \Bmodes\ modes to modulate the polarization signal;
    \item LiteBIRD-like in combination with CMB-S4-like, labeled ``\litesfour''.
\end{itemize}
The instrument characteristics of each simulated experiment are reported in Tab.~\ref{table: noise list}. In the following, when referring to individual experiments, we will omit the ``like'' suffix for brevity.

\begin{table}
\begin{center}
\begin{tabular}{ c|c|c|c }
\hline
 Experiment & Noise [$\mu$K-arcmin] & Beam size [arcmin] & Phase-shift $\beta$ [$^\circ$]\\
 \hline
 \hline
 LiteBIRD    &  2    &  30   &  0/10\\
 Planck      &  40   &  7    &  -   \\
 SO LAT      &  6.5  &  1.4  &  -   \\
 SO SAT      &  2    &  17   &  0/10\\
 S4-like LAT &  3    &  1.5  &  -   \\
 S4-like SAT &  1    &  30   &  0/10\\
 \hline
\end{tabular}
\caption{Instrumental specifications for each experiment: sensitivity to total intensity in $\mu$K-arcmin (linear polarization sensitivity is obtained by multiplying this value by a $\sqrt{2}$ factor, except for Planck which is 2 as only half the detectors are polarized); beam size FWHM in arcmin; non-ideal phase shift in degrees (for those experiments that will deploy - or we simulate will deploy - a HWP \cite{LiteBIRD:2020khw, SimonsObservatory:2018koc, Abazajian:2019eic}). The noise levels for SO SAT and LAT are reported for reference. They are coadded over the central 90 and 150 GHz channels, and the FWHM corresponds to the 145 GHz channel. For more details, see \cite{SimonsObservatory:2018koc, SimonsObservatory:2019qwx}. In this work, we make use of the official SO noise curves. 
}
\label{table: noise list}
\end{center}
\end{table}

As anticipated, for the sake of simplicity, we do not make use of the real data collected by Planck. We instead consider a simple white noise spectrum with $40$ $\mu$K-arcmin in temperature, and $80$ $\mu$K-arcmin in polarization ~\cite{Planck:2013oqw}, which reasonably reproduces the Planck performance. The noise spectrum for LiteBIRD is computed given the detector sensitivity and the angular resolution of LiteBIRD in the central frequency channels~\cite{LiteBIRD:2022cnt}. For SO~\cite{SimonsObservatory:2018koc}, we make use of the publicly available noise curves\footnote{SO noise curves are available at \url{https://github.com/simonsobs/so_noise_models}}. The latter also include residual contributions from component separation, which dominates the cosmological signal both at the very small scales probed by the SO LAT, as well as at the intermediate scales probed by the SO SAT. CMB-S4 noise curves were computed referring to the central channels for both SAT and LAT~\cite{CMB-S4:2020lpa}.

When adding \Vmodes-mode data in the analysis, the noise spectra of LiteBIRD, SO and CMB-S4 are further rescaled to account for the non-ideal phase shift, as detailed in Sec.~\ref{section: Sensitivity to V-modes}. In practice, the rescaled noise curves become:
\begin{equation} \label{eq: noise rescaling due to HWP}
    \begin{aligned}
\frac{\mathrm N_{\ell}^{\Emodes\Emodes}}{\mathrm B_\ell ^2} & \rightarrow \frac{1}{\mathrm B_\ell ^2} \: \frac{\mathrm N_{\ell}^{\Emodes\Emodes}}{\operatorname{Cos}^4(\beta /2)}\\
\frac{\mathrm N_{\ell}^{\Bmodes \Bmodes}}{\mathrm B_\ell ^2} & \rightarrow \frac{1}{\mathrm B_\ell ^2} \: \frac{\mathrm N_{\ell}^{\Bmodes \Bmodes}}{\operatorname{Cos}^4(\beta /2)}\\
\frac{\mathrm N_{\ell}^{\Vmodes \Vmodes}}{\mathrm B_\ell ^2} & \rightarrow \frac{1}{\mathrm B_\ell ^2} \: \frac{\mathrm N_{\ell}^{\Vmodes \Vmodes}}{\operatorname{Sin}^2(\beta)}.
    \end{aligned}
\end{equation}

Note that Planck/CMB-S4 and LiteBIRD experiments observe largely overlapping sky fractions. While LiteBIRD dominates over Planck in \Bmodes-mode sensitivity, the performance of the two experiments are different in T and \Emodes\ modes at different angular scales, depending on the respective noise level and angular resolution. A similar consideration applies to the combination ``\litesfour'' as far as the T and E-fields (``TE'' hereafter) are concerned. Therefore, for the ``TE'' fields in the ``\liteplanck'' and ``\litesfour'' cases, we use an effective inverse noise weighted curve:

\begin{equation}
\label{eq: effective noise}
    \mathrm{N}_\ell ^\mathrm{eff} =  \left[ \left( \frac{\mathrm{N}_\ell}{\mathrm{B} ^2 _\ell} \right)^{-1} _{\mathrm{X}} + \left( \frac{\mathrm{N}_\ell}{\mathrm{B} ^2 _\ell} \right)^{-1} _{\mathrm{LiteBIRD}} \right]^{-1},
\end{equation}
where $X=$ Planck , CMB-S4.
We can distinguish two regions where the two experiments perform differently: LiteBIRD dominates at larger scales, whereas Planck or CMB-S4 outperform LiteBIRD at smaller angular scales.\\
Differently, when Planck is combined with SO/CMB-S4, information from the former is only included in the sky fraction and/or at the angular scales not observed by SO/CMB-S4(LAT). As a result, in the ``\planckso/\plancksfour'' case, we make use of the individual noise curves for each experiment.\\

We employ an exact likelihood in harmonic space, rescaled by a $\ell$-independent sky fraction $f_\mathrm{sky}$ to account for the experiment-specific sky coverage.
\begin{equation} \label{eq: likelihood}
\begin{aligned}
-2 \ln\mathcal{L}_\ell = f_\mathrm{sky} (2 \ell+1) \left( \operatorname{Tr}\left[\mathbf{\hat{S}}_{\ell} \mathbf{S}_{\ell}^{-1}\right]+\ln \operatorname{det}\left[\mathbf{S}_{\ell}\right] \right) +\text { const. }
\end{aligned}
\end{equation}
where $\mathbf{\hat{S}}$ and $\mathbf{S}$ are matrices of the ``observed'' (i.e., simulated, in our case) $\hat{C}_\ell$ and theoretical $C_\ell$ power spectra, respectively (see, e.g., Sec. 2.2 of ~\cite{Gerbino:2019okg} for the definition of $\mathbf{\hat{S}}$ and $\mathbf{S}$). When \Vmodes\ modes are included in the likelihood analysis, the standard $\mathbf{\hat{S}}$ and $\mathbf{S}$ matrices are extended to include the $C_\ell^{\Vmodes \Vmodes}$. The sky fraction, range of angular scales and observed fields for each data combination\footnote{A note on the ``\planckso'' dataset observing TTTEEE (SO-LAT plus Planck): SO-LAT can at most cover 40\% of the sky, therefore we set Planck to observe the remaining 30\% of the sky to match an overall sky coverage of 70\% as in other cases. Similarly, for the ``\litesfour'' dataset on the combination observing TTTEEE (S4-LAT and LiteBIRD plus LiteBIRD alone): S4-LAT can at most cover 60\% of the sky, therefore we set LiteBIRD alone to observe the remaining 10\% of the sky to match the 70\% sky coverage.} are reported in Tab.~\ref{Table: experimental configuration}.

We simulate the observed spectra $\hat{C}_\ell$ as the sum of a reference set of CMB spectra obtained with the Boltzmann code \texttt{camb} \cite{2011ascl.soft02026L} from a fiducial set of cosmological parameters and the noise power spectrum in harmonic space: $\hat{C}_\ell=C_\ell^\mathrm{fid}+\mathrm N_\ell/\mathrm B_\ell^2$ (see, e.g., Sec. 3.3 of ~\cite{Gerbino:2019okg} for the extension of the definition of $\mathbf{\hat{S}}$ and $\mathbf{S}$ in Eq.~\ref{eq: likelihood} in presence of noise and beam smearing). We assume the fiducial model to be $\Lambda$CDM, i.e., vanishing \Vmodes-mode signal and vanishing primordial tensor modes. The cosmological parameters corresponding to the fiducial model are reported in Tab.~\ref{Table: fiducial values}. A major source of variance in the observation of primordial \Bmodes\ modes is the \Bmodes\Bmodes\ lensing contribution, which can be significantly reduced via appropriate de-lensing procedures. In order to quantify the effect of delensing in our results, we consider two cases: a fully lensed fiducial \Bmodes\Bmodes, i.e., no de-lensing, and a partly de-lensed fiducial \Bmodes\Bmodes. The two cases are obtained by properly rescaling the \Bmodes\Bmodes\ lensing contribution (in both the fiducial and the theory \Bmodes\Bmodes\ spectra) by a factor $A_{\rm L}=1$ (fully lensed) and $A_{\rm L}=0.3$ (70\% de-lensing). The GFE-modified (tensorial) \Bmodes\ modes are later summed to the $A_{\rm L}$-rescaled lensed \Bmodes\ modes. The underlying assumption, also used in Ref.~\cite{Lembo:2020ufn}, posits that GFE exclusively affects primordial \Bmodes\ modes, even though GFE and lensing should occur simultaneously as integrated effects along the line of sight. 
This assumption is valid for LiteBIRD and SO, see, e.g.,~\cite{Lembo:2020ufn}. However, for fourth-generation ground-based experiments such as CMB-S4, caution is advised regarding the interplay between lensing and GFE. In order to assess the validity of the assumption for CMB-S4, we compute the difference between the BB spectrum obtained as detailed above and the BB spectrum obtained by applying GFE to the sum of tensor and lensing BB spectra. We then compare this difference with the CMB-S4 noise curve. We find that, for reasonably small values of GFE parameters ($\beta_{E,V}^2\lesssim0.1$), the difference in the predicted CMB spectra between the two approaches is well below the noise contribution. Furthermore, other studies in the literature, such as Ref.~\cite{Namikawa:2024dgj}, have confirmed that lensed \Bmodes\ modes have a negligible impact on cosmic birefringence.

To better quantify the impact of a potential sensitivity to \Vmodes\ modes, we compare results obtained with and without the inclusion of \Vmodes\ modes in the datasets for each combination of experiments. We refer to the two cases as ``\Tmodes\Emodes\Bmodes\Vmodes'' and ``\Tmodes\Emodes\Bmodes'', respectively. In the ``\Tmodes\Emodes\Bmodes\Vmodes'' case, we assume the non-ideal phase shift to be $\beta=10^{\circ}$; as already noted, this choice is in agreement with the expected values of the phase shift from simulations. In addition, as shown in Fig.~\ref{fig:sensitivity vs beta}, this value allows concurrently for a relatively acceptable noise level in \Vstokes-Stokes parameter and negligible degradation of the noise level in \Qstokes/\Ustokes-Stokes. In the ``\Tmodes\Emodes\Bmodes'' case, we assume the phase shift to be $\beta=0$. 

\begin{table}[ht]
\begin{center}
\begin{tabular}{c|c|c|c|c }
Label & Experiment  & Multipoles & Fields & $f_{sky}$\\
\hline
\hline
\multirow{1}{*}{\litebird}    & LiteBIRD                & 2,1000    & TEB/TEBV & 0.7\\
\hline
\multirow{3}{*}{\liteplanck}  & LiteBIRD \& Planck      & 2,1500    & TE       & 0.7\\
                              & Planck                  & 1501,2500 & T        & 0.7\\
                              & LiteBIRD                & 2,1000    & B/BV     & 0.7\\
\hline
\multirow{4}{*}{\planckso}    & Planck                  & 2,50      & TE       & 0.7\\
                              & Planck                  & 51,2500   & TE       & 0.3\\
                              & SO LAT                  & 51,3000   & TE       & 0.4\\
                              & SO SAT                  & 50,300    & B/BV     & 0.1\\
\hline
\multirow{4}{*}{\plancksfour} & Planck                  & 2,50      & TE       & 0.7 \\
                              & Planck                  & 51,2500   & TE       & 0.1 \\
                              & S4-like LAT             & 51,3000   & TE       & 0.6 \\
                              & S4-like SAT             & 50,300    & B/BV     & 0.03\\
\hline
\multirow{5}{*}{\litesfour}   & LiteBIRD                & 2,50      & TEB/TEBV & 0.7 \\
                              & LiteBIRD                & 51,1000   & B/BV     & 0.7 \\
                              & S4-like LAT \& LiteBIRD & 51,3000   & TE       & 0.6 \\
                              & LiteBIRD                & 51,1000   & TE       & 0.1 \\
                              & S4-like SAT             & 50,300    & B/BV     & 0.03\\
\end{tabular}
\caption{Data combinations considered in this work. For each experiment, we report the observed fields (temperature T, linear polarization E and B, circular polarization V) and the corresponding range of angular scales (multipoles), as well as the sky fraction observed ($f_\mathrm{sky}$). Throughout the main text, we refer to each combination with the corresponding label in the first column.}
\label{Table: experimental configuration}
\end{center}
\end{table}

\begin{table}[ht]
\begin{center}
\renewcommand{\arraystretch}{1.1}
\begin{tabular}{c|c}
Cosmological parameter  & Fiducial value\\
 \hline
 \hline
$\Omega_{\mathrm{b}}h^2$   & 0.02237\\
$\Omega_{\mathrm{c}}h^2$   & 0.1200 \\
$100\theta_{\mathrm{MC}}$  & 1.04092\\
$\tau$                     & 0.0544 \\
$\ln(10^{10}\mathrm{A_s})$ & 3.044  \\
$n_{\mathrm{s}}$           & 0.9649 \\
\hdashline
$r$                        & 0      \\
$\beta^2 _E$               & 0      \\
$\beta^2 _V$               & 0      \\
\end{tabular}
\caption{Fiducial values of cosmological parameters for the reference cosmological model assumed to simulate the observed CMB spectra in this analysis. They are taken from the Planck best-fit values from \Tmodes\Tmodes{}, \Tmodes\Emodes{}, \Emodes\Emodes{} $+$ low\Emodes{} $+$ lensing analysis~\cite{Planck:2018vyg}.
}
\label{Table: fiducial values}
\end{center}
\end{table}
\section{Results} \label{section: Results}
In this section, we present and discuss the results obtained for the various datasets and cases introduced in the previous section. These results are also compared with the current bounds reported in literature~\cite{Lembo:2020ufn, BICEP:2021xfz}. We focus on a subset of cosmological parameters: the optical depth ($\tau$), the tensor-to-scalar ratio ($r$), and the GFE parameters ($\beta_E^2$ and $\beta_V^2$)\footnote{We do not report constraints on other parameters beyond these, as the impact of $\beta_E^2$ and $\beta_V^2$ on the rest of the parameter space is beyond the scope of this paper and, in any case, any effect is negligible. See, e.g., ~\cite{Caloni:2022kwp} for further details.}. The first two parameters are linked to the primary science goals of experiments observing large and intermediate-scale polarization, such as LiteBIRD, SO, and CMB-S4, while the latter two are the main focus of this study. The 68\% C.L. constraints on $\tau$ and the 95\% C.L. upper bounds on the remaining three parameters are in Tab. \ref{Table: all results} for both the fully lensed and partially de-lensed scenarios. Marginalized 1D and 2D posterior probabilities for $r$, $\tau$, $\beta^2_E$ and $\beta^2_V$ can be found in the Figures in Appendix \ref{section: Appendix for triangular plots}. We compare the results obtained with and without the inclusion of \Vmodes-mode data. In doing so, we expect to place more stringent limits on $\beta_E^2$ and $\beta_V^2$. In addition, we test whether the partial degradation of sensitivity to linear polarization resulting from a non-ideal phase shift affects the constraints on $\tau$ and $r$.\\ 
From Tab. \ref{Table: all results}, we can draw the following considerations:
\begin{itemize}
    \item as known, the constraints on $\tau$ and $r$ are driven by the ability of a given data combination to recover the large-scale \Emodes-mode signal and the large-to-intermediate \Bmodes-mode signal respectively. In Tab.~\ref{Table: all results}, compare, e.g., the improvement of the constraints on $\tau$ from the combinations including LiteBIRD with respect to those including Planck. Similarly, compare the lack of improvement of the constraints on $r$ from the ``\litebird'' dataset with respect to the ``\liteplanck'' combination.

    Ignoring for a moment the impact of \Vmodes-mode observations (but see the second bullet point below), the constraints on $\beta_E^2$ are mostly driven by the ability to reconstruct the TE and EE spectra at intermediate and small scales, since $\beta_E^2$ acts as an overall rescaling factor for the CMB spectra (see Eqs.~\ref{eq: cmb power spectra}). Indeed, by comparing the first rows of each Dataset box in Tab.~\ref{Table: all results}, we note a steady improvement of the constraints on $\beta_E^2$ when considering dataset which are more and more sensitive to small-scale \Emodes\ (compare, e.g., ``\litebird'' and ``\liteplanck''; also, note the lack of improvement when comparing ``\plancksfour'' with ``\litesfour'', as CMB-S4 dominates the sensitivity to \Emodes\ in both data combinations). 

    The constraints on $\beta_V^2$ are mostly driven by the sensitivity of each dataset to \Bmodes\ modes, both via the overall rescaling of the amplitude (first term on the RHS in Eq.~\ref{eq: cmb power spectra}) and via the mixing with \Emodes\ modes (second term on the RHS)\footnote{
    The same mixing term appears on the RHS of the equation for $C_\ell^{\Emodes \Emodes}$. However, this term is negligible with respect to the first one appearing on the RHS of the same equation, since the amplitude of $\Tilde{C}_\ell^{\Bmodes \Bmodes}$ (as well as the amplitude of the $\Tilde{C}_\ell^{\Emodes \Emodes}$ weighted by the Wigner coefficients) is significantly lower than the amplitude of $\Tilde{C}_\ell^{\Emodes \Emodes}$. Therefore, \Emodes-mode observations do not play a significant role in constraining $\beta_V^2$ compared to \Bmodes-mode observations;
    }.  
    
    \item the addition of \Vmodes-mode data tightens the constraints on $\beta^2_E$ by roughly an order of magnitude but has no effect on $\beta^2_V$ (compare the two rows corresponding to each data label in Tab.~\ref{Table: all results}; see also Appendix~\ref{section: Appendix for triangular plots}). Indeed, $C_{\ell}^{\Vmodes \Vmodes}$ in Eq.~\ref{eq: cmb power spectra} is solely affected by $\beta^2_E$ that acts as an overall rescaling factor of a mixing of $\Tilde{C}_\ell^{\Emodes \Emodes}$ and $\Tilde{C}_\ell^{\Bmodes \Bmodes}$. The inclusion of \Vmodes\ modes, with the corresponding rescaling of the noise spectra as per Eq.~\ref{eq: noise rescaling due to HWP} and Fig.~\ref{fig:sensitivity vs beta}, does not degrade the constraints on $r$ and $\tau$. This is one of the main findings of this work. This stands in contrast to the assertion made in~\cite{Lembo:2020ufn} regarding the necessity of improving linear polarization for increased accuracy on GFE parameters. The inclusion of \Vmodes-mode data proves to be an additional, significant source of constraining power, beyond what is provided by linear polarization alone. The sensitivity (i.e. the low \Vmodes\Vmodes\ noise) is enhanced by the non-ideal phase shift of the HWP (as shown in Fig.~\ref{fig:sensitivity vs beta}), where a relatively conservative value of $\beta=10^\circ$ was employed;
    
    \item by reducing the contribution of signal variance in the observation of primordial \Bmodes\ modes, the de-lensing procedure results in a factor of two improvement of the constraints on $r$ and $\beta^2_V$ (compare rows with the same data label in the top and bottom panels in Tab.~\ref{Table: all results}). No improvement is seen for $\tau$ and $\beta^2_E$, as could be visually verified by comparing the left (no de-lensing) with the right (70\% de-lensing) panel of triangular plots in Appendix~\ref{section: Appendix for triangular plots}. This confirms that the constraining power on these parameters comes essentially from TE and EE spectra rather than from BB (in addition to VV for $\beta_E^2$).\\
    Improvements on $r$ and $\beta^2_V$ are expected, since they directly impact the $C_\ell^{\Bmodes \Bmodes}$ power spectrum as rescaling factors. In Eq.~\ref{eq: cmb power spectra}, we recognize an attenuating factor $(\beta^2_V + \beta^2_E)/4\pi$ in front of the unmodified $\Tilde{C}_\ell^{\Bmodes \Bmodes}$. In the second term on the RHS, the amplitude of the combination of unmodified \Emodes\Emodes\ and \Bmodes\Bmodes\ spectra - with the $\Tilde{C}_\ell^{\Bmodes \Bmodes}$ spectrum being subdominant compared to the unmodified $\Tilde{C}_\ell^{\Emodes \Emodes}$ - is regulated by $\beta^2_V$. This requires $\beta^2_V$ to suppress the power coming from the \Emodes\Emodes\ spectrum, thus restoring the expected amplitude of the observed tensorial \Bmodes\ modes ($r=0$).\\
    Also, de-lensing does not help improve the constraining power on the optical depth $\tau$. In principle, information on $\tau$ can be also extracted from measurements of the reionization peak in the BB spectrum. However, the strong degeneracy between $r$ and $\tau$ in that region as well as the residual lensing variance dominating also at low-$\ell$s for vanishingly small values of $r$ severely limit the information content that can be extracted from BB to improve the constraints on $\tau$.
    As a result, the constraining power for this parameter comes essentially from \Emodes\ modes only;
    
    \item when comparing the constraints on $r$ from ``\planckso'' and ``\plancksfour'' in the fully lensed case (top panel in Tab.~\ref{Table: all results}), we note that the latter dataset provides - somehow unexpectedly - a looser bound on $r$ than the former dataset (see also left panels of Figs.~\ref{fig: triangular planck so} and~\ref{fig: triangular planck s4}). This feature can be attributed to the fact that the two experiments driving the constraints on $r$ in the two data combinations - namely, SO and CMB-S4 - are both cosmic variance limited in \Bmodes\Bmodes\ in the fully lensed scenario. With the instrumental noise being subdominant, the ability to jointly constrain the parameters affecting $C_\ell^{\Bmodes \Bmodes}$ is solely determined by volume effects in the parameter space explored in the MCMC analysis. More in detail, at first order, we can assume that the amplitude of $C_\ell^{\Bmodes \Bmodes}$ is determined by a combination of $r$, $\beta_E^2$ and $\beta_V^2$ (see Eqs.~\ref{eq: cmb power spectra}). When switching from ``\planckso'' to ``\plancksfour'' data combination, the enhanced sensitivity of the latter to \Emodes\ modes leads to more stringent limits on $\beta_E^2$, thus allowing for a larger excursion of $r$ in parameter space. For the same reason, the constraints on $\beta_V^2$ from the two data combinations are virtually unchanged. In this case, the volume effect responsible for the degradation of the constraints on $r$ is suppressed for $\beta_V^2$ by the extra contribution of the \Emodes-\Bmodes\ mixing term on the RHS in the equation for $C_\ell^{\Bmodes \Bmodes}$. 
    
    Conversely, for partial de-lensing (bottom panel of Tab.~\ref{Table: all results}; see also right panels of Figs.~\ref{fig: triangular planck so} and~\ref{fig: triangular planck s4}), we observe the expected improvement of the constraint on $r$ when moving from ``\planckso'' to ``\plancksfour''. While ``\plancksfour'' remains cosmic variance limited, the impact of instrumental noise in ``\planckso'' cannot be neglected and washes out the volume effect described above.
\end{itemize}
Finally, in Tab.~\ref{table: final results} we quote again the 95\% C.L. upper limits for the five analyzed cases combining TEBV modes, and compare them with current bounds. We note that current limits in the literature are obtained without a de-lensing procedure applied to data. Nevertheless, we report forecasts for both the fully lensed and the partially de-lensed scenarios, to also assess the benefit of de-lensing.
The current upper limit on $r$ is taken from the analysis of BICEP2, Keck Array, BICEP3, WMAP, and Planck data, marginalizing over a given model of Galactic dust and synchrotron contaminations~\cite{BICEP:2021xfz}. As already mentioned, no de-lensing is applied to the data. Future experiments targeting $r$ will see considerable improvement in the constraints by applying suitable de-lensing techniques (see, e.g.,~\cite{CMB-S4:2020lpa,Belkner:2023duz,Wolz:2023lzb}). Our forecasts qualitatively confirm this expectation.
Present bounds on $\beta^2_{E,V}$ are taken from the analysis of~\cite{Lembo:2020ufn} and were derived using temperature and linear polarization anisotropies from the Planck legacy release~\cite{Planck:2019nip} and BICEP2/Keck 2015~\cite{BICEP2:2018kqh}
. The inclusion of current \Vmodes-mode observations does not help improve the constraints on the GFE parameters~\cite{Lembo:2020ufn}. Forecasted sensitivity to $\beta_V^2$ is dominated by the sensitivity of upcoming experiments to \Bmodes\ modes, with de-lensing playing a crucial role. As far as $\beta_E^2$ is concerned, while de-lensing has minimal impact on the forecasted sensitivity, the inclusion of \Vmodes\ modes from future experiments significantly improves the constraints. 
Our forecast shows that more stringent constraints ranging from 1 to 3 orders of magnitude can be achieved, depending on the considered parameter and dataset.

\begin{table}[ht]
\centering
\renewcommand{\arraystretch}{1.3} 

\begin{tabular}{ c|c|c|c|c|c }
Dataset $\left(A_{\rm L}=1\right)$ & Fields & $\tau$ & $\beta^2_E \;[10^{-2}]$ & $\beta^2_V\;[10^{-3}]$ & $r \;[10^{-3}]$ \\
\hline \hline

\multirow{2}{*}{\litebird}   & TEB  & $0.0543^{+0.0019}_{-0.0021}$ & $< 9.7 $ & $< 1.1 $ & $< 0.4 $\\
                             & TEBV & $0.0544 \pm 0.0020$          & $< 0.5 $ & $< 1.1 $ & $< 0.4 $\\\hline
\multirow{2}{*}{\liteplanck} & TEB  & $0.0544^{+0.0019}_{-0.0021}$ & $< 8.4 $ & $< 1.1 $ & $< 0.4 $\\
                             & TEBV & $0.0544^{+0.0018}_{-0.0021}$ & $< 0.5 $ & $< 1.1 $ & $< 0.4 $\\\hline
\multirow{2}{*}{\planckso}   & TEB  & $0.0547 \pm 0.0040$          & $< 6.6 $ & $< 4.2 $ & $< 4.4 $\\
                             & TEBV & $0.0546 \pm 0.0040$          & $< 2.9 $ & $< 4.3 $ & $< 4.4 $\\\hline
\multirow{2}{*}{\plancksfour}& TEB  & $0.0548^{+0.0034}_{-0.0038}$ & $< 5.1 $ & $< 4.5 $ & $< 5.7 $\\
                             & TEBV & $0.0547^{+0.0034}_{-0.0038}$ & $< 1.0 $ & $< 4.4 $ & $< 5.0 $\\\hline
\multirow{2}{*}{\litesfour}  & TEB  & $0.0544^{+0.0017}_{-0.0020}$ & $< 5.1 $ & $< 0.9 $ & $< 0.4 $\\
                             & TEBV & $0.0544 \pm 0.0019$          & $< 0.5 $ & $< 0.8 $ & $< 0.4 $
\end{tabular}

\vspace{1cm}

\begin{tabular}{ c|c|c|c|c|c }
Dataset $\left(A_{\rm L}=0.3\right)$ & Fields & $\tau$ & $\beta^2_E \;[10^{-2}]$ & $\beta^2_V\;[10^{-3}]$ & $r \;[10^{-3}]$ \\
\hline \hline

\multirow{2}{*}{\litebird}   & TEB  & $0.0543^{+0.0018}_{-0.0021}$ & $< 9.8 $ & $< 0.6 $ & $< 0.2 $\\
                             & TEBV & $0.0544^{+0.0018}_{-0.0021}$ & $< 0.5 $ & $< 0.6 $ & $< 0.2 $\\\hline
\multirow{2}{*}{\liteplanck} & TEB  & $0.0543^{+0.0018}_{-0.0021}$ & $< 8.7 $ & $< 0.6 $ & $< 0.2 $\\
                             & TEBV & $0.0544 \pm 0.0019$          & $< 0.5 $ & $< 0.6 $ & $< 0.2 $\\\hline
\multirow{2}{*}{\planckso}   & TEB  & $0.0549^{+0.0037}_{-0.0041}$ & $< 6.6 $ & $< 2.5 $ & $< 2.5 $\\
                             & TEBV & $0.0549^{+0.0038}_{-0.0043}$ & $< 3.0 $ & $< 2.6 $ & $< 2.5 $\\\hline
\multirow{2}{*}{\plancksfour}& TEB  & $0.0549 \pm 0.0036$          & $< 5.2 $ & $< 1.8 $ & $< 2.0 $\\
                             & TEBV & $0.0547 \pm 0.0037$          & $< 1.0 $ & $< 1.8 $ & $< 2.0 $\\\hline
\multirow{2}{*}{\litesfour}  & TEB  & $0.0545 \pm 0.0019$          & $< 5.3 $ & $< 0.5 $ & $< 0.2 $\\
                             & TEBV & $0.0545^{+0.0017}_{-0.0020}$ & $< 0.5 $ & $< 0.5 $ & $< 0.2 $
\end{tabular}
\caption{Constraints on a subset of cosmological parameters sampled in this analysis: 68\% C.L. constraints for the optical depth ($\tau$), and 95\% C.L. upper limits for the tensor-to-scalar ratio ($r$) and the GFE parameters ($\beta^2_E$, $\beta^2_V$). The results are shown assuming a lensing factor $A_{\rm L}=1$ (no de-lensing, top table) and $A_{\rm L}=0.3$ (70\% de-lensing efficiency, bottom table), for different data combinations (\litebird, \liteplanck, \planckso, \plancksfour{}, and \litesfour). For each data combination, we compare the results obtained with and without the inclusion of \Vmodes-mode observations.}
\label{Table: all results}
\end{table}

\begin{table}[ht!]
\centering
\renewcommand{\arraystretch}{1.1}
\begin{tabular}{c||c|c|c||c|c|c} 
\multirow{2}{*}{Dataset} & \multicolumn{3}{c||}{95\% C.L. $\left(A_{\rm L}=1\right)$} & \multicolumn{3}{c}{95\% C.L. $\left(A_{\rm L}=0.3\right)$}\\
  & $\beta^2 _E \;[10^{-2}]$ & $\beta^2_V \;[10^{-3}]$ & $r\; [10^{-3}]$ & $\beta^2 _E \;[10^{-2}]$ & $\beta^2 _V[10^{-3}]$ & $r\;[10^{-3}]$\\ [0.5ex]
   
 \hline\hline
 \litebird       & $< 0.5 $ & $< 1.1 $ & $< 0.4 $ & $< 0.5 $ & $< 0.6 $ & $< 0.2 $\\
 \liteplanck     & $< 0.5 $ & $< 1.1 $ & $< 0.4 $ & $< 0.5 $ & $< 0.6 $ & $< 0.2 $\\
 \planckso       & $< 2.9 $ & $< 4.3 $ & $< 4.4 $ & $< 3.0 $ & $< 2.6 $ & $< 2.5 $\\
 \plancksfour    & $< 1.0 $ & $< 4.4 $ & $< 5.0 $ & $< 1.0 $ & $< 1.8 $ & $< 2.0 $\\
 \litesfour      & $< 0.5 $ & $< 0.8 $ & $< 0.4 $ & $< 0.5 $ & $< 0.5 $ & $< 0.2 $\\
 \hline
 Current bounds  & $< 14  $ & $< 30 $   & $< 36 $ &          &          &         \\ 
\end{tabular}

\vspace{0cm}

\caption{95\% C.L upper limits on the tensor-to-scalar ratio and GFE parameters ($\beta_E^2$ and $\beta_V^2$) for the different data combinations considered in this analysis. These limits are obtained from the analysis of \Tmodes, \Emodes, \Bmodes\ and \Vmodes\ modes, with a lensing factor $A_{\rm L}=1$ (second, third and fourth columns) and $A_\mathrm{L}=0.3$ (last three columns). In the last row, bounds on the same parameters from current data~\cite{Lembo:2020ufn, BICEP:2021xfz} are also reported to ease the comparison, where no de-lensing procedures are included.
}
\label{table: final results}
\end{table}

\section{Conclusions} \label{section: Conclusions}
In this study, we investigated how the deployment of a realistic half-wave plate (HWP) inducing a non-ideal (i.e., $\neq \pi$) phase shift, may affect the noise characteristics of CMB experiments, while providing sensitivity to the presence of a non-vanishing \Vmodes-mode signal. We then forecasted the expected performance of new-generation CMB experiments, noting that our analysis did not consider foregrounds.

Deviations from the ideal phase shift modify the noise properties of a CMB experiment deploying a HWP by degrading the sensitivity to \Qstokes{}/\Ustokes\ polarization. This could potentially spoil the ability to reach the primary science goals of future CMB experiments, such as a CVL determination of the optical depth $\tau$ and a tight constraint on/high-significance detection of the tensor-to-scalar ratio $r$. However, at the same time, deviations from the ideal performance of a HWP allow for the experiment to become sensitive to possibly non-vanishing \Vmodes\ modes in the observed signal. This work showed how even a significant deviation from the ideal phase shift (of the order of $10^\circ{}$) can open a window to the investigation of beyond-the-standard-model (BSM) cosmological scenarios while having no impact on the sensitivity to $\tau$ and $r$. 

As a test case, we focused on the phenomenology of a class of models leading to BSM electromagnetic effects on the CMB, the Generalized Faraday Effect (GFE)~\cite{Lembo:2020ufn}. The GFE is responsible for \Emodes-\Bmodes\ mixing as well as for sourcing circular polarization from the conversion of linear CMB polarization. These effects are modeled via two phenomenological parameters - $\beta_E^2$ and $\beta_V^2$ - which modify the shape of the CMB spectra as they emerge from the last scattering surface (Eqs.~\ref{eq: cmb power spectra}). We forecasted the sensitivity to these parameters from currently running and future CMB experiments which will or might deploy a HWP. At the same time, we checked that the sensitivity to $\tau$ and $r$ expected from these experiments is not degraded. We considered different combinations of simulated data from a LiteBIRD-like satellite experiment (\textit{L}), from the ground-based Simons Observatory (\textit{SO}) and a modified version (i.e., with the inclusion of a HWP) of a ground-based CMB-S4-like experiment (\textit{S4})\footnote{We remind the reader that the baseline configuration of CMB-S4 does not include the HWP. In this work, we considered the possibility to deploy a HWP as a way to test the performance of a future ground-based experiment with the expected sensitivity and angular resolution of CMB-S4.}. We also included information from Planck (\textit{P}) observations where relevant. 

By comparing the results with current limits in the literature~\cite{Lembo:2020ufn}, we forecasted a 1-to-3 order-of-magnitude improvement of the constraints on $\beta_V^2$ and $\beta_E^2$, depending on the dataset, with the most stringent limits achieved with the ``\litesfour'' combination (see Tab.~\ref{table: final results}). 

The possibility to include observations of \Vmodes\ modes, as enabled by the deployment of a realistic HWP, is key to obtain dramatic improvements in the constraints on $\beta_E^2$, ranging from a factor of a few for the combinations ``\planckso'' and ``\plancksfour'' to an order of magnitude and more for the data combinations including LiteBIRD (see Tab.~\ref{Table: all results}). Better sensitivity to intermediate-to-small scale \Emodes\ modes is also an important driver of the improved constraints on $\beta_E^2$, as can be noted by comparing ``\litebird'', ``\liteplanck'', ``\planckso''/``\plancksfour''. Constraints on $\beta_V^2$ are mostly insensitive to the inclusion of \Vmodes\ modes in the analysis. Instead, they are driven by improved sensitivity to \Bmodes\ modes. The constraints improve roughly by a factor of two when partial de-lensing (obtained as a 70\% suppression of lensing power in the simulated \Bmodes\Bmodes\ data) is applied in the analysis (compare the top and bottom panels in Tab.~\ref{Table: all results}).
The stringent constraints on $\beta_V^2$ and $\beta_E^2$ will provide valuable insights on the fundamental physics governing the evolution of the Universe, particularly from the CMB last scattering to the present epoch. The GFE parameters can be mapped onto the elements of an effective ``cosmic susceptibility tensor'', which describes the optical properties of the Universe seen as the medium through which CMB photons propagate. The mapping between $\beta_V^2$ and $\beta_E^2$ and the elements of the susceptibility tensor can be obtained once a specific physical model responsible for the GFE is specified. 
As such, constraints on $\beta_V^2$ and $\beta_E^2$ shed light on the nature of the GFE and its potential origins, such as magnetic fields~\cite{De:2014qza, Melrose:1991eu, Lemarchand:2018lfy, Ejlli:2016avx, Ejlli:2018ucq, Cooray:2003} or other BSM phenomena~\cite{Alexander:2019sqb, Motie:2011az, Sawyer:2014maa, Li:2024fxy, Caloni:2022kwp} that could have influenced the polarization state of the CMB.
The methodology outlined in this work can be also applied to models capable of generating parity-breaking spectra (e.g. Ref.~\cite{Caloni:2022kwp}). 
In conclusion, the results presented in this study indicate that the possibility of including observations of \Vmodes\ modes in the cosmological analysis is a key tool to investigate more deeply specific aspects of fundamental physics which are otherwise less accessible, without spoiling the constraining power to other primary science targets.
This emphasizes the necessity of fully exploit the potential of upcoming CMB experiments, including the extraction of \Vmodes-mode data, which will provide unprecedented insights on the fundamental physics of the early Universe.

\acknowledgments
 We thank Alessandro Gruppuso for useful discussions during the preparation of the manuscript and for feedback on the final version of the paper.
 We acknowledge the financial support from the INFN InDark initiative and from the COSMOS network (www.cosmosnet.it) through the ASI (Italian Space Agency) Grants 2016-24-H.0 and 2016-24-H.1-2018, as well as 2020-9-HH.0 (participation in LiteBIRD, phase A). We acknowledge the use of \texttt{camb}~\cite{2011ascl.soft02026L}, \texttt{CosmoMC}~\cite{PhysRevD.66.103511}, \texttt{GetDist} \cite{Lewis:2019xzd}, and the use of computing facilities at CINECA. MG is funded by the European Union (ERC, RELiCS, project number 101116027) and by the PRIN (Progetti di ricerca di Rilevante Interesse Nazionale) number 2022WJ9J33. SG acknowledges support from the Horizon 2020 ERC Starting Grant (Grant agreement No 849169) and from STFC and UKRI (grant numbers ST/W002892/1 and ST/X006360/1). This is not an official SO Collaboration paper.

\bibliographystyle{JHEP}
\bibliography{bibliography,Planck_bib}
\appendix
\section{Complete set of triangular plots}\label{section: Appendix for triangular plots}
\begin{figure}[htbp]
    \centering
    \begin{subfigure}[b]{0.49\textwidth}
        \centering
        \includegraphics[width=\textwidth]{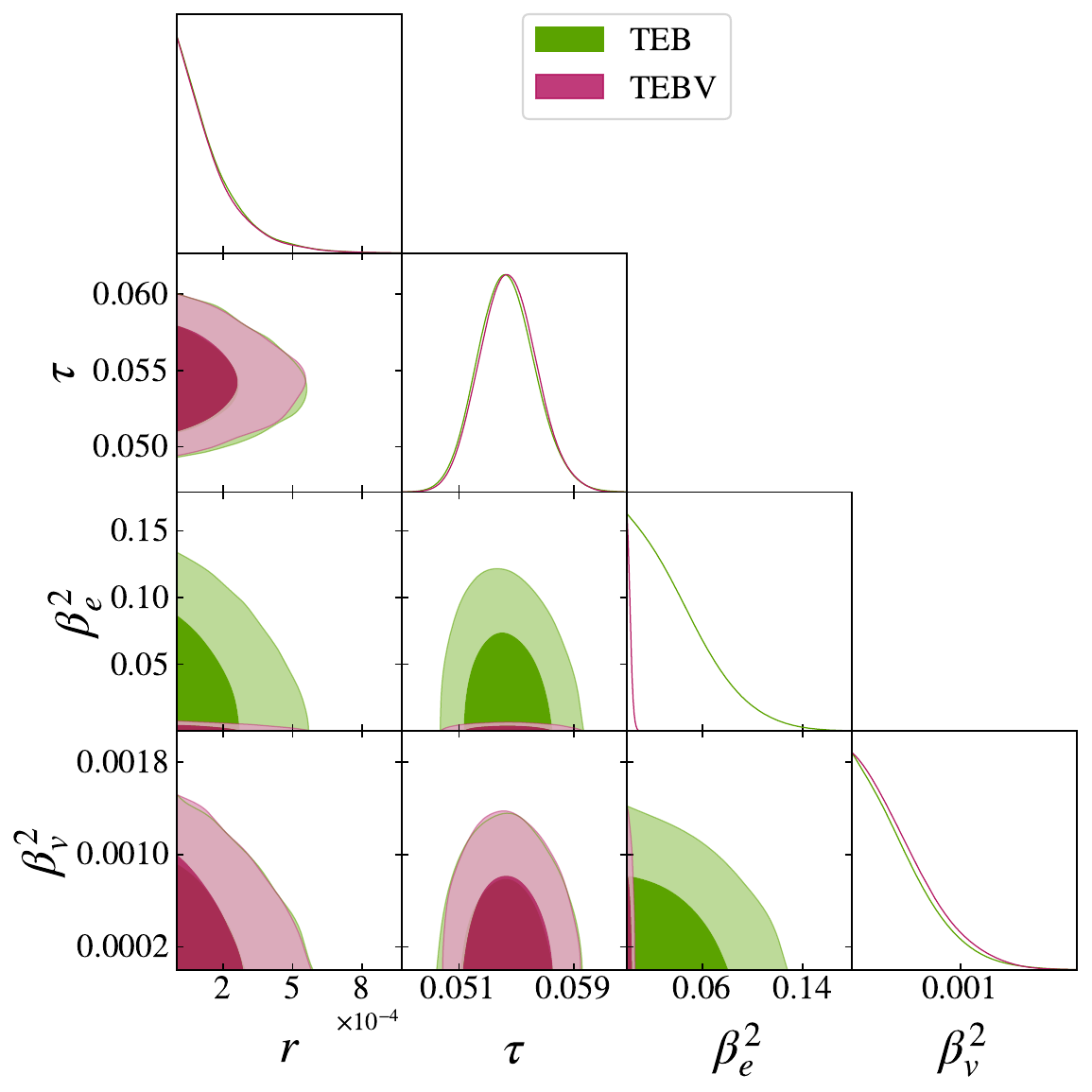}
    \end{subfigure}
    \hfill
    \begin{subfigure}[b]{0.49\textwidth}
        \centering
        \includegraphics[width=\textwidth]{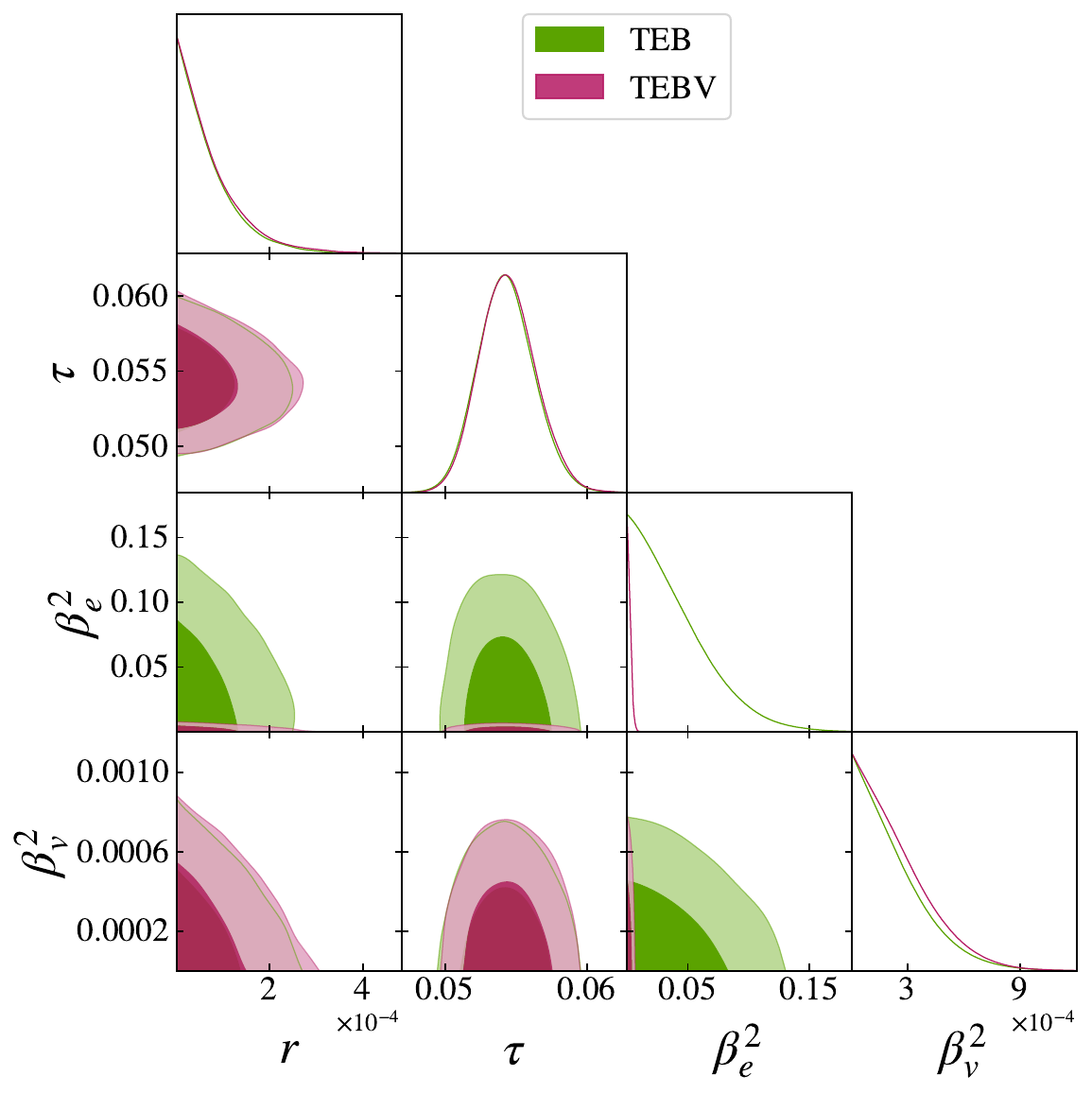}
    \end{subfigure}
    \caption{
    One and two-dimensional posterior probability distributions for \litebird{} dataset (see Tab.~\ref{Table: experimental configuration} for details). We report a subset of the total parameter space ($\Lambda$CDM model $+ \: r+\beta^2 _E + \beta^2 _V$) showing TTTEEEBB in green and TTTEEEBBVV in magenta. The left panel omits de-lensing procedures ($A_{\rm L}=1$), while the right panel employs 70\% de-lensing ($A_{\rm L}=0.3$).
    }
    \label{fig: triangular litebird}
\end{figure}

\begin{figure}[htbp]
    \centering
    \begin{subfigure}[b]{0.49\textwidth}
        \centering
        \includegraphics[width=\textwidth]{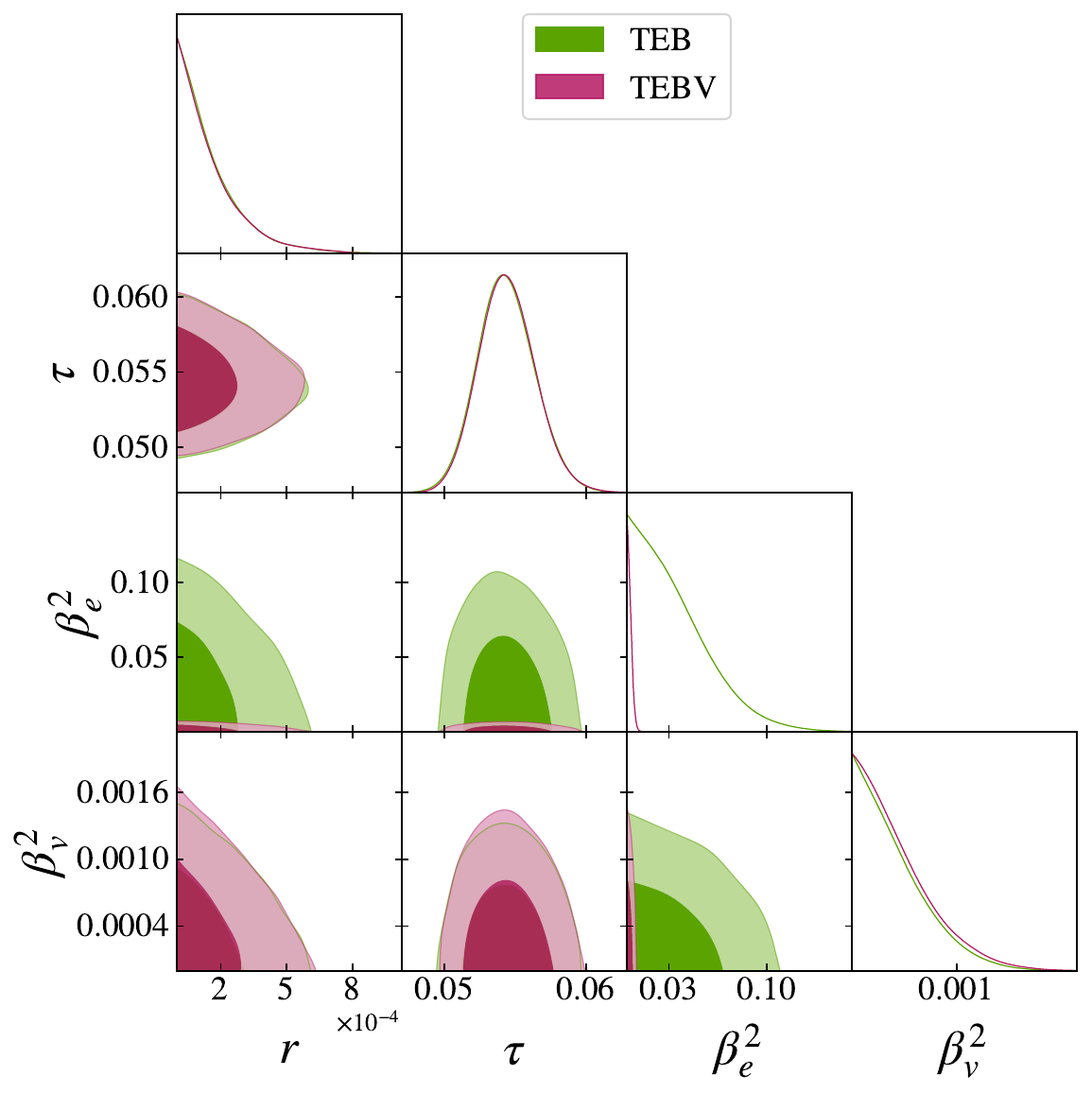}
    \end{subfigure}
    \hfill
    \begin{subfigure}[b]{0.49\textwidth}
        \centering
        \includegraphics[width=\textwidth]{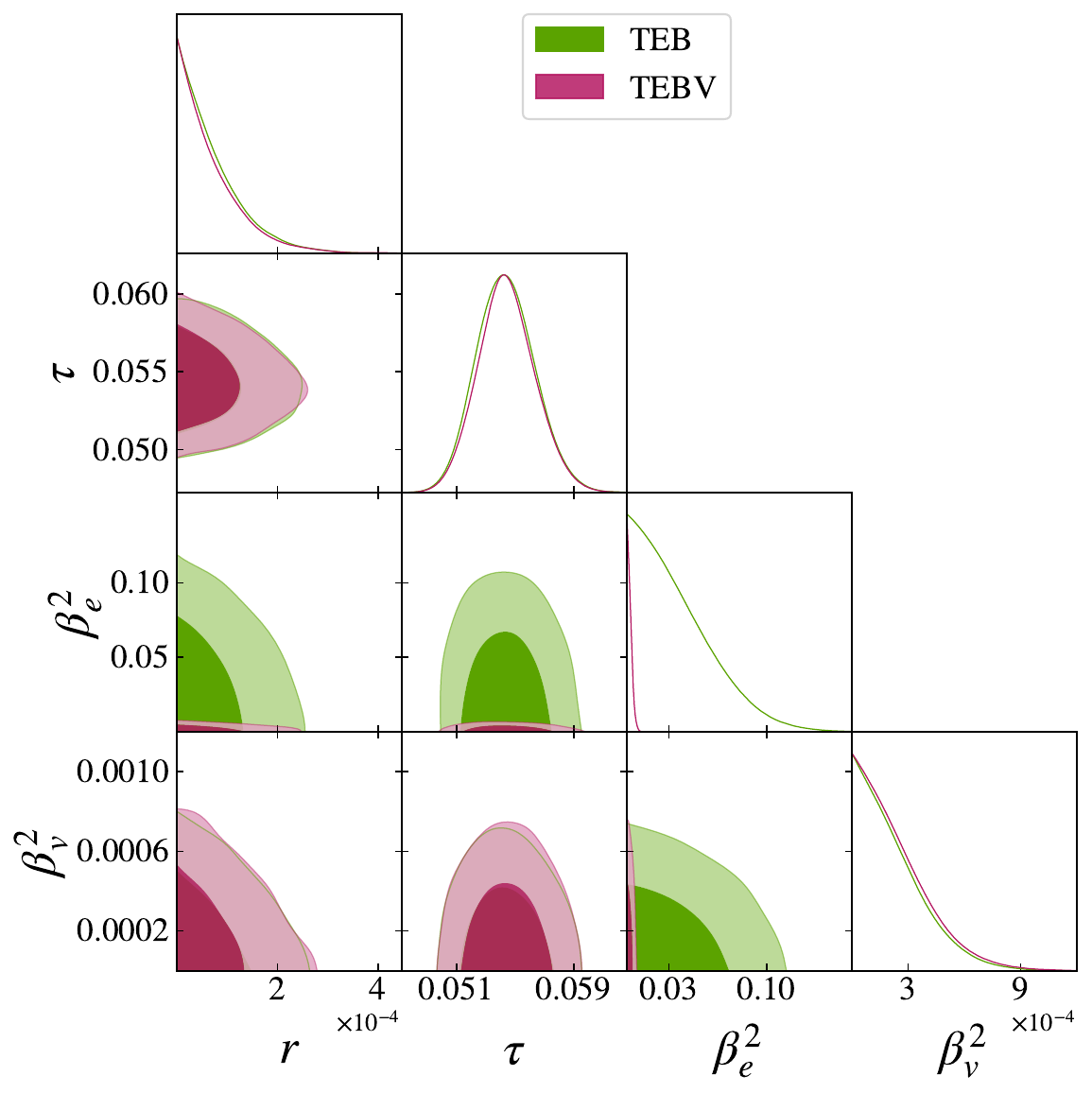}
    \end{subfigure}
    \caption{
    One and two-dimensional posterior probability distributions for \liteplanck{} dataset (see Tab.~\ref{Table: experimental configuration} for details). We report a subset of the total parameter space ($\Lambda$CDM model $+ \: r+\beta^2 _E + \beta^2 _V$) showing TTTEEEBB in green and TTTEEEBBVV in magenta. The left panel omits de-lensing procedures ($A_{\rm L}=1$), while the right panel employs 70\% de-lensing ($A_{\rm L}=0.3$).
    }
    \label{fig: triangular litebird planck}
\end{figure}

\begin{figure}[htbp]
    \centering
    \begin{subfigure}[b]{0.49\textwidth}
        \centering
        \includegraphics[width=\textwidth]{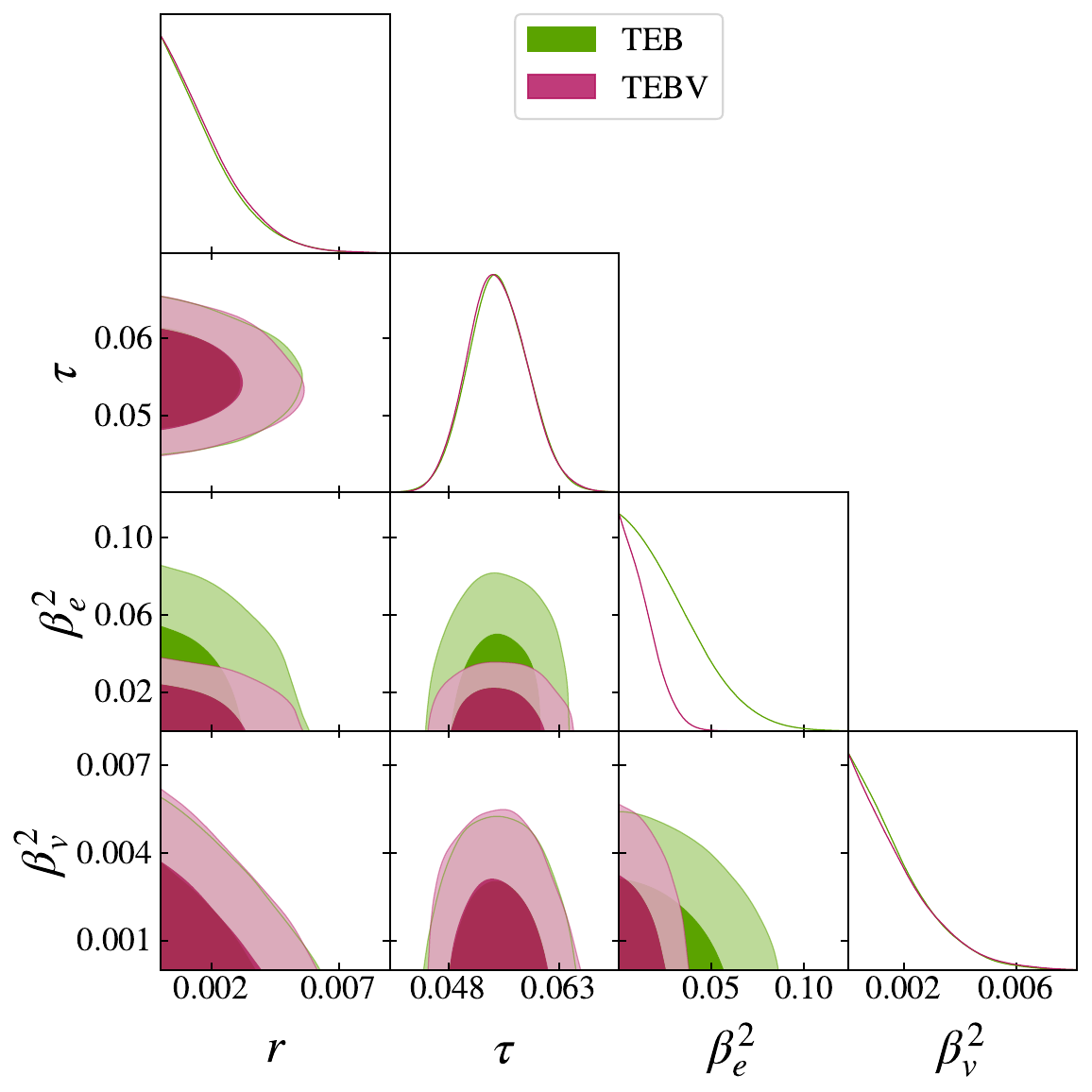}
    \end{subfigure}
    \hfill
    \begin{subfigure}[b]{0.49\textwidth}
        \centering
        \includegraphics[width=\textwidth]{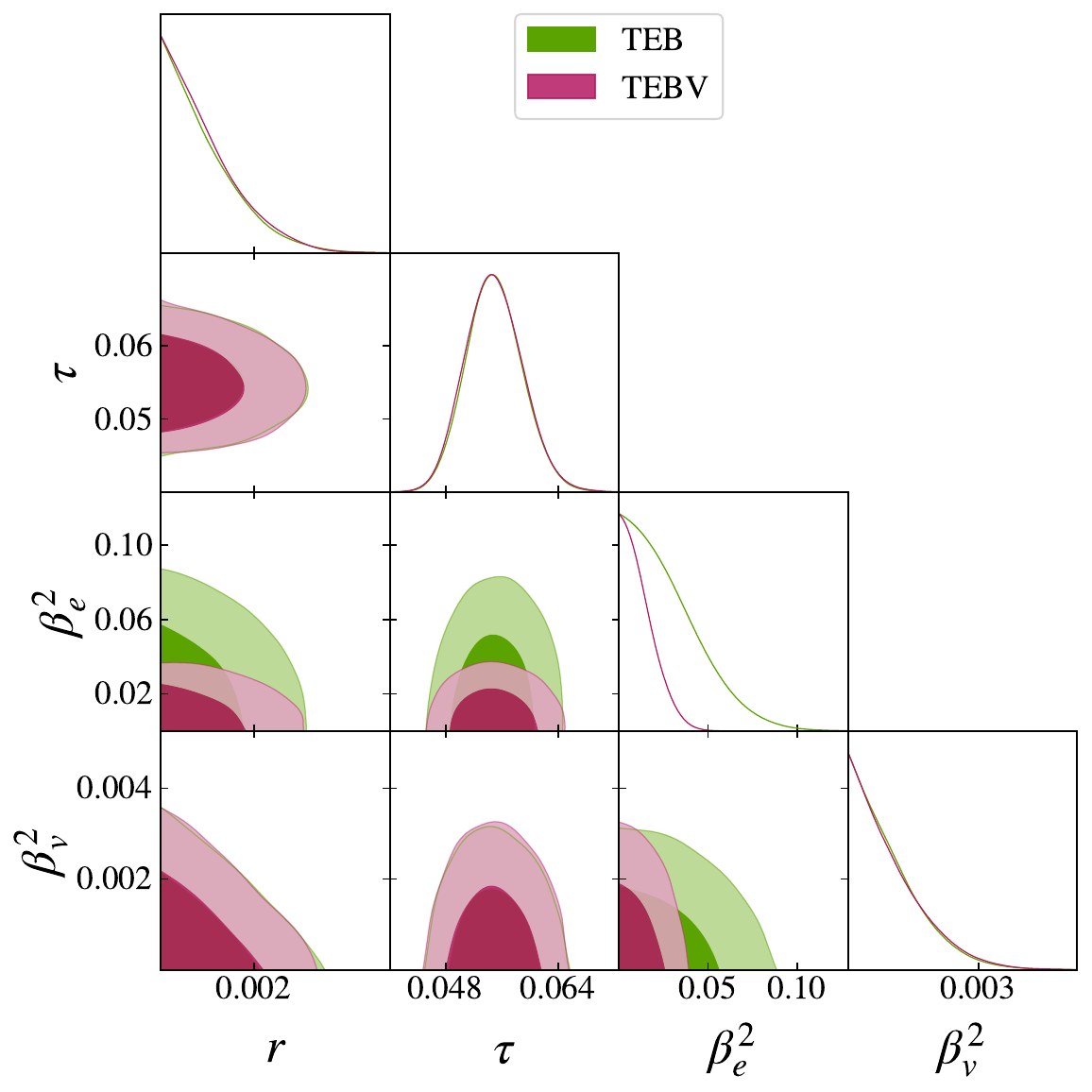}
    \end{subfigure}
    \caption{
    One and two-dimensional posterior probability distributions for \planckso\ dataset (see Tab.~\ref{Table: experimental configuration} for details). We report a subset of the total parameter space ($\Lambda$CDM model $+ \: r+\beta^2 _E + \beta^2 _V$) showing TTTEEEBB in green and TTTEEEBBVV in magenta. The left panel omits de-lensing procedures ($A_{\rm L}=1$), while the right panel employs 70\% de-lensing ($A_{\rm L}=0.3$).}
    \label{fig: triangular planck so}
\end{figure}

\begin{figure}[htbp]
    \centering
    \begin{subfigure}[b]{0.49\textwidth}
        \centering
        \includegraphics[width=\textwidth]{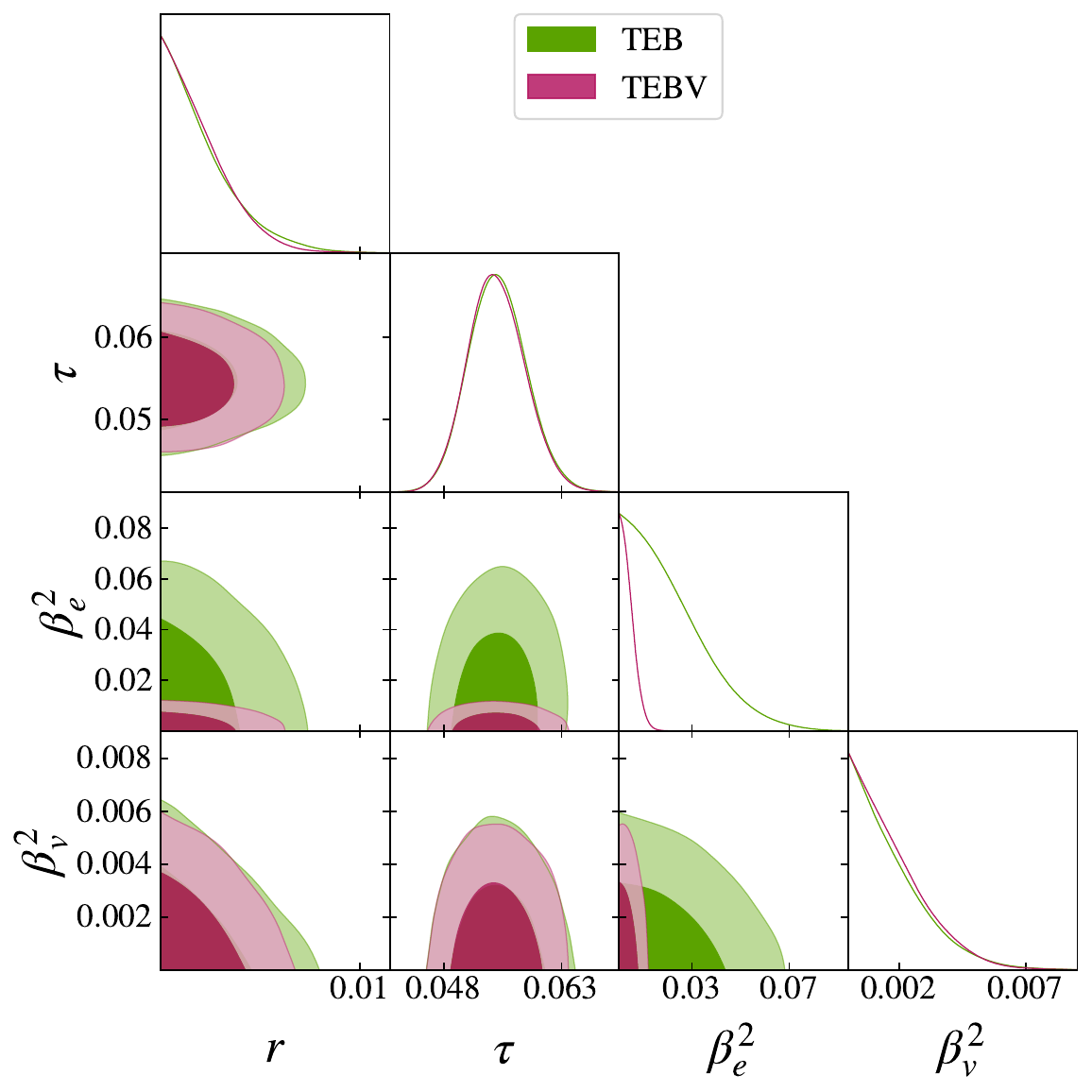}
    \end{subfigure}
    \hfill
    \begin{subfigure}[b]{0.49\textwidth}
        \centering
        \includegraphics[width=\textwidth]{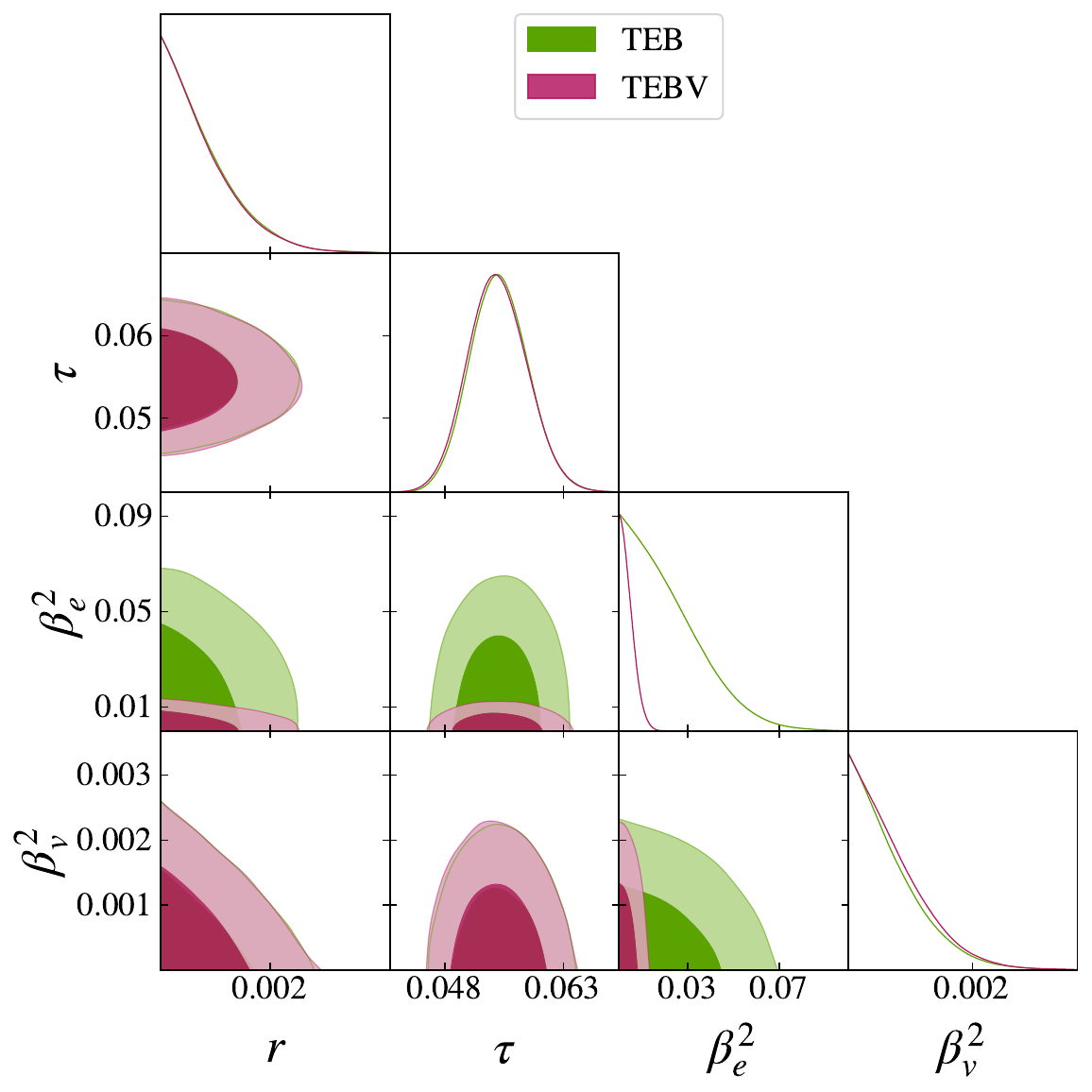}
    \end{subfigure}
    \caption{
    One and two-dimensional posterior probability distributions for \plancksfour\ dataset (see Tab.~\ref{Table: experimental configuration} for details). We report a subset of the total parameter space ($\Lambda$CDM model $+ \: r+\beta^2 _E + \beta^2 _V$) showing TTTEEEBB in green and TTTEEEBBVV in magenta. The left panel omits de-lensing procedures ($A_{\rm L}=1$), while the right panel employs 70\% de-lensing ($A_{\rm L}=0.3$).
    }
    \label{fig: triangular planck s4}
\end{figure}

\begin{figure}[htbp]
    \centering
    \begin{subfigure}[b]{0.49\textwidth}
        \centering
        \includegraphics[width=\textwidth]{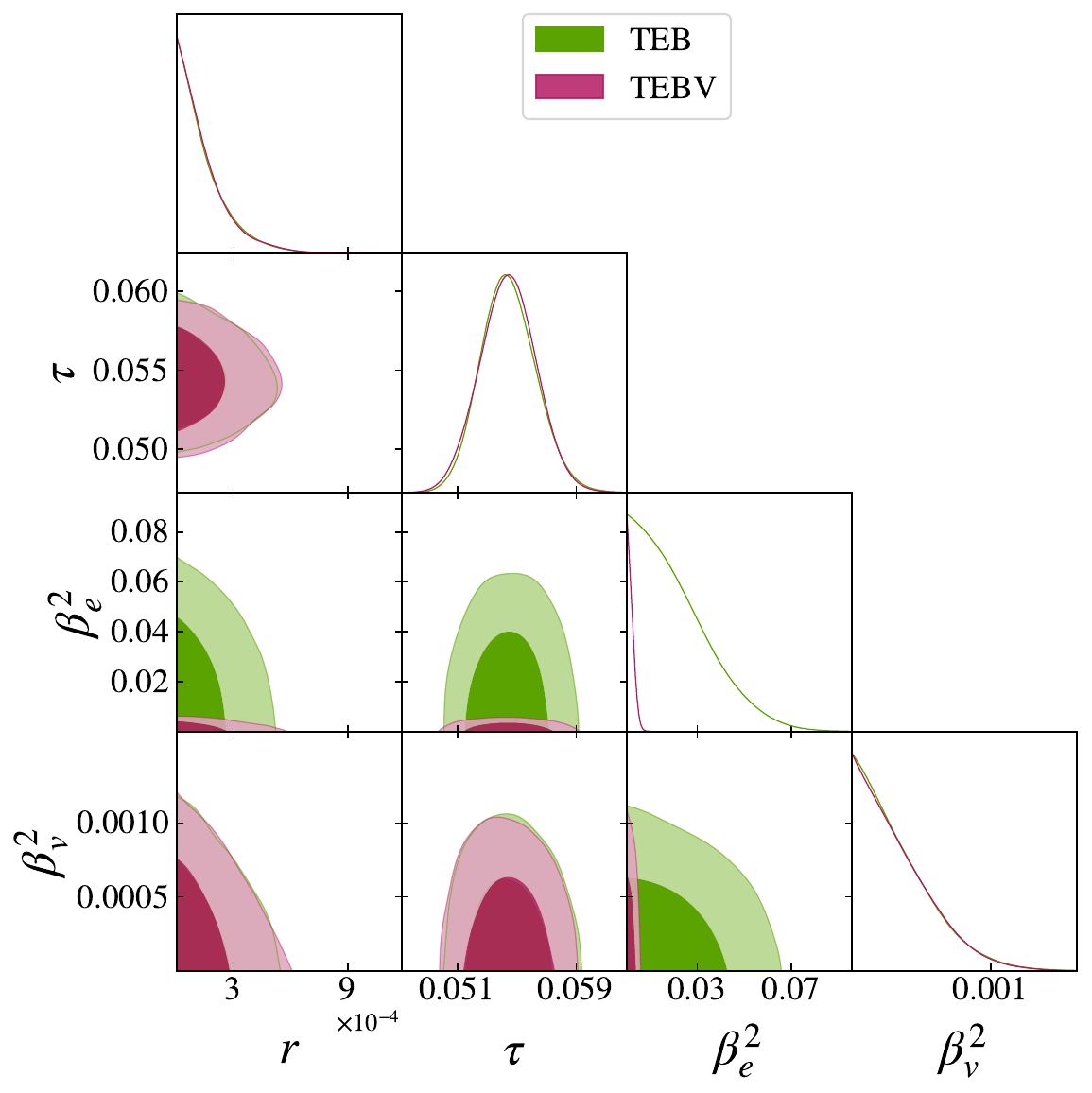}
    \end{subfigure}
    \hfill
    \begin{subfigure}[b]{0.49\textwidth}
        \centering
        \includegraphics[width=\textwidth]{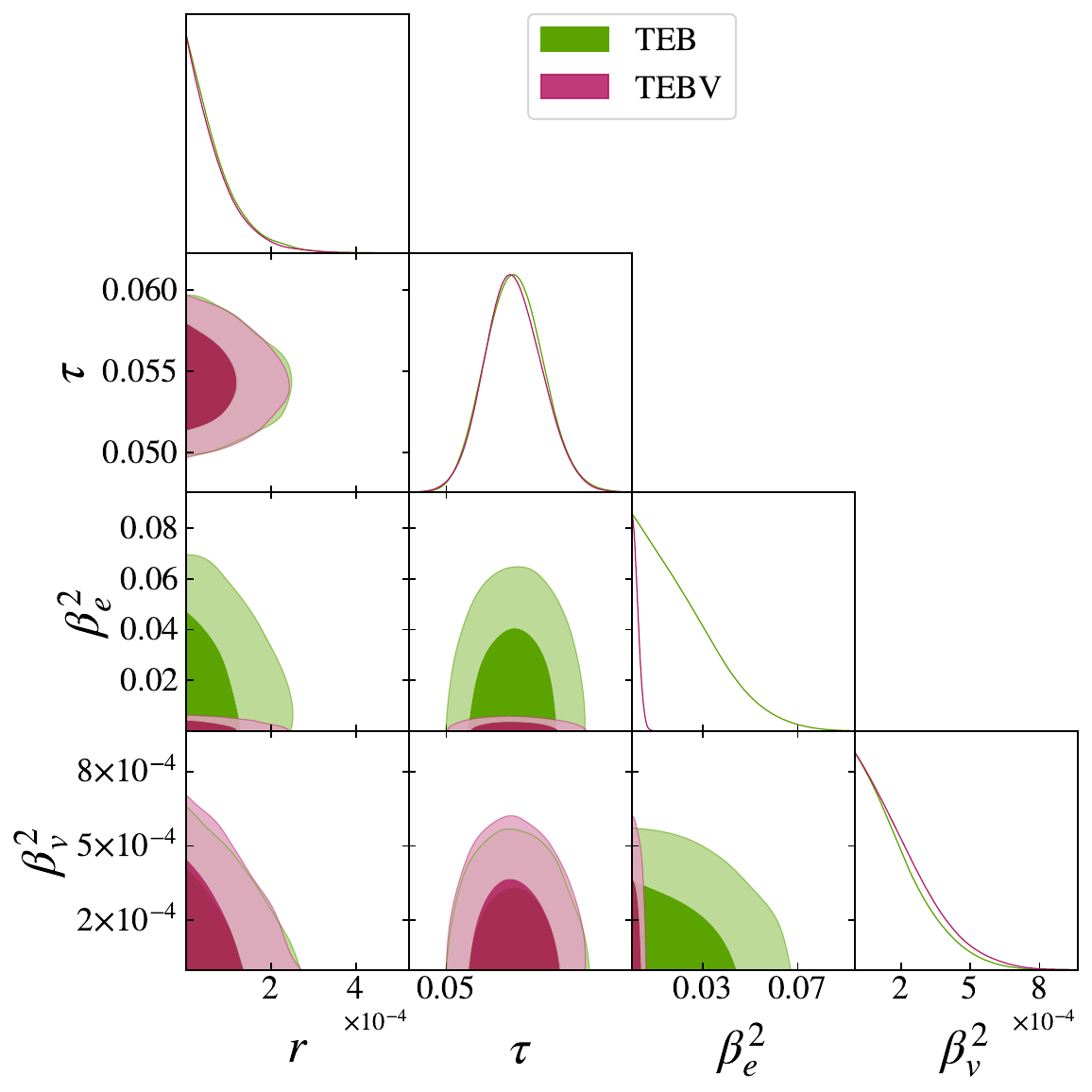}
    \end{subfigure}
    \caption{
    One and two-dimensional posterior probability distributions for \litesfour{} dataset (see Tab.~\ref{Table: experimental configuration} for details). We report a subset of the total parameter space ($\Lambda$CDM model $+ \: r+\beta^2 _E + \beta^2 _V$) showing TTTEEEBB in green and TTTEEEBBVV in magenta. The left panel omits de-lensing procedures ($A_{\rm L}=1$), while the right panel employs 70\% de-lensing ($A_{\rm L}=0.3$).
    }
    \label{fig: triangular litebird s4}
\end{figure}

We show here the triangular plots for the five cases analyzed, displaying 1D and 2D posteriors for the parameters we consider relevant in our analysis, i.e. $r$, $\tau$, $\beta^2_E$ and $\beta^2_V$. The general behaviour of all plots looks similar, the inclusion of \Vmodes-mode polarization substantially enhances our constraining power improving the bound on $\beta^2_E$. Furthermore, the addition of the de-lensing procedure is effective in reducing the upper limits of $r$ and $\beta^2_V$ (see the right panel of Figs.~\ref{fig: triangular litebird},~\ref{fig: triangular litebird planck},~\ref{fig: triangular planck so},~\ref{fig: triangular planck s4},~\ref{fig: triangular litebird s4}). In contrast, as expected, bounds on $\tau$ were found to be unaffected by both the de-lensing and the addition of \Vmodes-mode data.

\end{document}